\documentclass[prd,nofootinbib,preprint,superscriptaddress,twocolumn,10pt]{revtex4}
\pdfoutput=1
\usepackage{amsmath,amssymb}
\usepackage{epsfig}
\usepackage{graphicx}
\usepackage[usenames,dvipsnames]{color}
\usepackage{subfigure}
\usepackage{slashed}
\usepackage{comment}
\usepackage[colorlinks,citecolor=blue]{hyperref}
\usepackage{color}

\usepackage{multirow}
\usepackage{booktabs}

\begin{document}

\title{Self-interacting dark matter with observable $\Delta N_{\rm eff}$}
%in a $U(1)_D$ framework}
	
	%%%%%%%%%   Authors   %%%%%%%%%%%%		
	\author{Debasish Borah}
    \email{dborah@iitg.ac.in}
    \affiliation{Department of Physics, Indian Institute of Technology Guwahati, Assam 781039, India}
    \affiliation{Pittsburgh Particle Physics, Astrophysics, and Cosmology Center, Department of Physics and Astronomy, University of Pittsburgh, Pittsburgh, PA 15260, USA}
    
    \author{Satyabrata Mahapatra}
	\email{satyabrata@skku.edu}
	\affiliation{Department of Physics and Institute of Basic Science,
		Sungkyunkwan University, Suwon 16419, Korea}
		
	\author{Narendra Sahu}
	\email{nsahu@phy.iith.ac.in}
	\affiliation{Department of Physics, Indian Institute of Technology Hyderabad, Kandi, Sangareddy 502285, Telangana, India}
	
	\author{Vicky Singh Thounaojam}
    \email{ph22resch01004@iith.ac.in}
	\affiliation{Department of Physics, Indian Institute of Technology Hyderabad, Kandi, Sangareddy 502285, Telangana, India}
    
	\begin{abstract}

We propose a GeV-scale self-interacting dark matter (SIDM) candidate within a dark $U(1)_D$ gauged extension of the Standard Model (SM), addressing small-scale structure issues in $\Lambda$CDM while predicting an observable contribution to $\Delta N_{\rm eff}$ in the form of dark radiation. The model introduces a fermionic DM candidate $\chi$ and a scalar $\phi$, both charged under an unbroken $U(1)_D$ gauge symmetry. The self-interactions of $\chi$ are mediated by a light vector boson $X^\mu$, whose mass is generated via the Stueckelberg mechanism. The relic abundance of $\chi$ is determined by thermal freeze-out through annihilations into $X^\mu$, supplemented by a non-thermal component from the late decay of $\phi$. Crucially, $\phi$ decays after the Big Bang Nucleosynthesis (BBN) but before the Cosmic Microwave Background (CMB) epoch, producing additional $\chi$ and a dark radiation species ($\nu_S$). This late-time production compensates for thermal underabundance due to efficient annihilation into light mediators, while remaining consistent with structure formation constraints. The accompanying dark radiation yields a detectable $\Delta N_{\rm eff}$, compatible with Planck 2018 bounds and within reach of next-generation experiments such as SPT-3G, CMB-S4, and CMB-HD.

	\end{abstract}	
	\maketitle
	%\flushbottom
	
\noindent
\section{Introduction}\label{sec:Introduction}
Self-interacting dark matter (SIDM) has emerged as a compelling alternative to collisionless cold dark matter (CDM), addressing small-scale structure anomalies such as the too-big-to-fail, missing satellite, and core-cusp problems that afflict CDM models~\cite{Spergel:1999mh, Tulin:2017ara, Bullock:2017xww}. The key distinction lies in SIDM's self-interaction cross-section, parametrized as $\sigma/m \sim 1\,{\rm cm}^2/{\rm g} \approx 2 \times 10^{-24}\,{\rm cm}^2/{\rm GeV}$ \cite{Buckley:2009in, Feng:2009hw, Feng:2009mn, Loeb:2010gj}, for elastic scattering which can be achieved via light mediators. This interaction generates velocity-dependent effects that reconcile small-scale observations while preserving CDM's success at large scales \cite{Buckley:2009in, Feng:2009hw, Feng:2009mn, Loeb:2010gj, Bringmann:2016din, Kaplinghat:2015aga, Aarssen:2012fx, Tulin:2013teo}. 

Nevertheless, when dark matter (DM) couples to a light mediator with sizeable coupling, the annihilation rates become significantly enhanced, often leading to an underabundance of thermal relic for DM in the low-mass regime (typically a few GeV). Although the light thermal DM scenario ($M_{\rm DM} \lesssim \mathcal{O}(10~\rm GeV)$) has garnered considerable attention, especially due to the relatively weak constraints from direct detection experiments~\cite{LUX-ZEPLIN:2022qhg}, obtaining the correct relic density in this regime remains challenging~\cite{Borah:2022ask,Borah:2023sal,Mahapatra:2023oyh,Borah:2024wos,Adhikary:2024btd}. For WIMP-like interactions, avoiding overclosure of the Universe generally imposes a lower bound on the DM mass around a few GeV~\cite{Lee:1977ua, Kolb:1985nn}. However, by introducing light mediators or additional fields, it is possible to realize viable light thermal DM scenarios, as demonstrated in several recent works~\cite{Pospelov:2007mp, DAgnolo:2015ujb, Berlin:2017ftj, DAgnolo:2020mpt, Herms:2022nhd, Jaramillo:2022mos, Borah:2024yow}. In the context of SIDM, the efficient annihilation into light mediators typically depletes the thermal relic density. Therefore, a hybrid production mechanism that combines both thermal and non-thermal contributions can be an attractive route to achieving the observed DM abundance~\cite{Dutta:2021wbn, Borah:2021pet, Borah:2021rbx, Borah:2021qmi}.

\begin{figure}[h]
	\centering
    \includegraphics[width=0.35\textwidth]{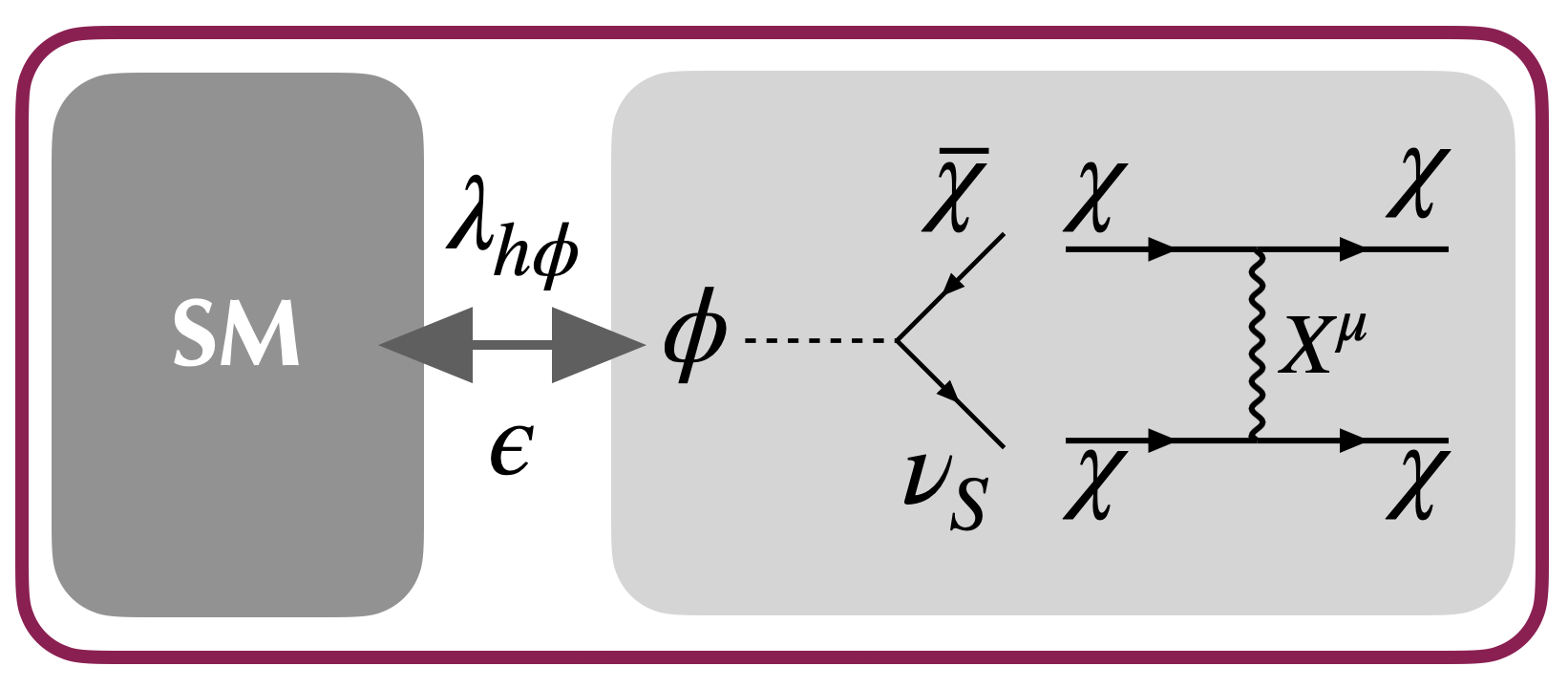}
	\caption{Schematic of the proposed setup: dark matter ($\chi$) with self-interactions mediated by a light dark gauge boson $X^\mu$. Late time production of DM $\chi$ and DR ($\nu_S$) from a scalar $\phi$ decay compensating for thermal underabundance of DM while also generating observable DR.}
	\label{fig:portal}
\end{figure}

In this work, we explore a gauged $U(1)_D$ framework that accommodates a GeV-scale SIDM candidate, along with observable deviations in the effective number of relativistic species, $\Delta N_{\rm eff}$. The dark sector is comprised of a fermionic DM candidate $\chi$, a dark radiation state $\nu_S$, a singlet scalar $\phi$, and a vector mediator $X^\mu$. The self-interactions of $\chi$ are mediated by the light vector boson $X^\mu$, while the $U(1)_D$ symmetry remains unbroken, ensuring the stability of DM. The mass of $X^\mu$ considered to be generated via the Stueckelberg mechanism \cite{Stueckelberg:1938hvi} for simplicity, although the Higgs mechanism is also viable. A central challenge in the GeV-scale SIDM scenario is that the efficient annihilation of $\chi$ into light mediators often results in an under-abundant relic density. In our model, this deficit is remedied by the late decay of the singlet scalar $\phi$, which not only replenishes the DM population but also produces additional dark radiation ($\nu_S$). The latter contributes to $\Delta N_{\rm eff}$, potentially yielding observable signatures in CMB experiments. As current CMB measurements constrain $N_{\rm eff} = 2.99^{+0.34}_{-0.33}$ (Planck 2018 \cite{Planck:2018vyg}), with future experiments (SPT-3G\cite{SPT-3G:2019sok}, CMB-S4 \cite{Abazajian:2019eic}, CMB-HD \cite{CMB-HD:2022bsz}) poised to probe deviations below $\Delta N_{\rm eff} \sim 0.06$, this setup offers a complementary cosmological probe of the SIDM, while being consistent with constraints from structure formation and direct detection of DM. While we consider $\nu_S$ to be an additional light fermionic degree of freedom, $\phi$ can also decay into $\chi$ and active neutrinos $\nu_L$ at late epochs leading to enhanced $\Delta N_{\rm eff}$. Such an interaction $\phi \bar{\chi} \nu_L$ can be facilitated by active-sterile mixing below the electroweak symmetry breaking scale, similar to type-I seesaw model of neutrino mass. Such an UV completion does not require any new light degrees of freedom. In the present work, we remain agnostic about such UV completion and consider $\nu_S$ to be a new light degree of freedom.

This paper is organized as follows: in Section~\ref{sec::model} we present the details the gauged $U(1)_D$ model under consideration. Section~\ref{sec::relic} investigates the parameter space where the correct DM relic density and observable $\Delta N_{\rm eff}$ are simultaneously achieved while Section~\ref{sec::Neff} discusses the contributions to $\Delta N_{\rm eff}$ and the relevant constraints. Direct detection prospects of DM are studied in Section~\ref{sec::dd}, while Section~\ref{sec::additional} examines $\phi$ decay kinematics and DM free-streaming. We finally conclude in Section~\ref{sec::conclusion}.

\section{The Model}\label{sec::model}

The model extends the Standard Model (SM) gauge group $SU(3)_c \times SU(2)_L \times U(1)_Y$ with an additional $U(1)_D$ symmetry. The dark sector (DS) consists of: a Dirac fermion $\chi$ (DM candidate), a complex scalar $\phi$, and a sterile neutrino $\nu_S$. All new fields are singlets under the SM gauge group. The $U(1)_D$ charges are assigned such that $\chi$ and $\phi$ carry equal charge while $\nu_S$ is neutral. Crucially, $\phi$ does not acquire a vacuum expectation value (vev), preventing $\chi$-$\nu_S$ mixing that would destabilize the DM. A schematic illustration of the dark sector is presented in Fig.~\ref{fig:portal}. In our setup, the mass of $X^\mu$ arises due to Stueckelberg mechanism, as discussed in Appendix~\ref{app:stueckelberg}.

% This model is based on $SU(3)_C \times SU(2)_L \times U(1)_Y \times U(1)_D$ gauge symmetry. In addition to the SM particles we introduce a singlet Dirac dark matter $\chi$, a complex singlet scalar $\phi$ and the sterile neutrinos $\nu_S$, which are all singlet under SM gauge group. A pictorial presentation is shown in Fig.~\ref{fig:portal}.Under the newly introduced $U(1)_D$ symmetry, $\chi$ and $\phi$ have the same charge, while $\nu_S$ transforms trivially. However, $\phi$ is not allowed to acquire a vacuum expectation value (vev), as this would result in mixing between $\chi$ and $\nu_S$, leading to the instability of $\chi$. 

The relavant lagrangian is given by:
\begin{equation}
\begin{split}
    \mathcal{L} \supset & ~(D^\mu \phi)^\dagger (D_\mu \phi) + \mathcal{L_{Z-X}} - m_{\chi}\overline{\chi}\chi \\
    & - (y\phi \overline{\chi}\nu_S + \text{h.c.}) - V(\phi,\Phi)
\end{split}
\end{equation}
where the scalar potential is given by:
\begin{equation}\label{eqn::potential}
\begin{split}
    V(\phi,\Phi) = & -\mu_h^2 \Phi^\dagger \Phi + \lambda_h (\Phi^\dagger \Phi)^2 \\
    & + \mu_\phi^2 |\phi|^2 + \lambda_\phi |\phi|^4 + \lambda_{h\phi} |\phi|^2 \Phi^\dagger \Phi.
\end{split}
\end{equation}
Here $\Phi$ denotes the SM Higgs doublet with $h$ being the physical scalar after electroweak symmetry breaking. $\mathcal{L_{Z-X}}$, which describes the the dynamics of the gauge fields in the interaction basis ($\tilde{W}^3_\mu$, $\tilde{B}_\mu$, $\tilde{X}_\mu$) reads:
\begin{equation}\label{eqn::Z-X}
	\begin{split}
		\mathcal{L_{Z-X}} \supset & -\frac{1}{4} \Tilde{B}_{\mu\nu}\Tilde{B}^{\mu\nu}-\frac{1}{4} \Tilde{W}_{3\mu\nu}\Tilde{W}^{3\mu\nu} \\
		& -\frac{1}{4} \Tilde{X}_{\mu\nu}\Tilde{X}^{\mu\nu}-\frac{\epsilon}{2} \Tilde{X}_{\mu\nu}\Tilde{B}^{\mu\nu} \\
        & + j_{SM}^\mu \Tilde{B}_\mu + j_{\Tilde{X}}^\mu \Tilde{X}_\mu.
	\end{split}
\end{equation}
where $\tilde{V}_{\mu\nu}$ denotes the field strength tensors and $\epsilon$ parametrizes kinetic mixing between dark gauge boson field strength $\tilde{X}_{\mu\nu}$ and the hypercharge field strength $\tilde{B}_{\mu\nu}$. In Eq.~\eqref{eqn::Z-X}, the SM electroweak neutral current is given by $j^\mu_{\rm SM}=\{j^\mu_{\rm em},\,j^\mu_Z\}$, and the dark current is $j^\mu_X=g_d\,\overline{\chi}\gamma^\mu \chi$. To write the above Lagrangian in the canonical form
one needs to diagonalize the kinetic terms of the gauge bosons.

% Here, we introduce $\Tilde{X}_{\mu}$ as the new gauge boson of $U(1)_D$ and $\Tilde{X}_{\mu\nu}$ is the corresponding field strength tensor. $\epsilon$ is a parameter that controls the strength of the mixing between $\Tilde{X}_{\mu\nu}$ and $\Tilde{B}_{\mu\nu}$.
%In Eqn.~\ref{eqn::Z-X}, $j^\mu_{SM}=\{j^\mu_{em},j^\mu_Z\}$ and $j^\mu_X=g_d \overline{\chi}\gamma^\mu \chi$ represent the SM current and the dark current respectively. % In the new basis, the Lagrangian, including the Stueckelberg mass term of $\Tilde{X}^\mu$ (Eqn.~\ref{eq:Stueckelberg_term}),  becomes

After electroweak symmetry breaking, with $\Phi = (0, (v+h)/\sqrt{2})^T$,  where $h$ is the physical Higgs boson and $v$ is the vev,  
%Following electroweak symmetry breaking,
%of the Standard Model (SM),
we rotate to the intermediate basis as:
\begin{equation}
        \begin{pmatrix}
            \Tilde{W}_\mu^3 \\
            \Tilde{B}_\mu \\
            \Tilde{X_\mu}
        \end{pmatrix}\\
         =  \begin{pmatrix}
            cw & sw & 0\\
            -sw & cw & 0 \\
            0 & 0 & 1
            \end{pmatrix}
         \begin{pmatrix}
            \hat{Z}_\mu\\
            \hat{A}_\mu \\
            \hat{X_\mu}
        \end{pmatrix}
\end{equation}
where $cw=\rm cos\theta_W $ and $sw=\rm sin\theta_W$ with $\rm \theta_W$ being the Weinberg angle. 
Thus the Lagrangian in the new basis, including the Stueckelberg mass term for $\tilde{X}_\mu$, becomes \cite{Kors:2004dx}
\begin{equation}
	\begin{split}
		\mathcal{L_{Z-X}} \supset & -\frac{1}{4} \hat{A}_{\mu\nu}\hat{A}^{\mu\nu}-\frac{1}{4} \hat{Z}_{\mu\nu}\hat{Z}^{\mu\nu}-\frac{1}{4} \hat{X}_{\mu\nu}\hat{X}^{\mu\nu} \\
		& -\frac{\epsilon}{2} ~cw~ \hat{X}_{\mu\nu}\hat{A}^{\mu\nu}-\frac{\epsilon}{2} ~sw~ \hat{X}_{\mu\nu}\hat{Z}^{\mu\nu} \\
        & + j_{\rm em}^\mu \hat{A}_\mu +j_{Z}^\mu \hat{Z}_\mu + j_{X}^\mu \hat{X}_\mu\\
        & + \frac{1}{2}m_X^2~\hat{X_\mu}\hat{X^\mu} + \frac{1}{2}m_Z^2~\hat{Z_\mu}\hat{Z^\mu}.
	\end{split}
    \label{eq:Lagrangian}
\end{equation}
Here, $j^\mu_{Z}=-\overline{f}\gamma^\mu(g_L^f P
_L +g_R^f P_R) f$ and $j^\mu_{\rm em}=-e Q_f \overline{f} \gamma^\mu f$.
The mixing terms proportional to $\epsilon$ can be eliminated by a field redefinition as:
% Note that, in the 2nd line of Eqn.~\ref{eq:Lagrangian}, the $\epsilon$ terms lead to a mixing between $\hat{X}^\mu$ with $\hat{A}^\mu$ and $\hat{Z}^\mu$. This can be avoided by making a further redefinition of the fields as 
\begin{equation}
    \begin{pmatrix}
        \hat{A}_\mu \\
        \hat{Z}_\mu \\
        \hat{X}_\mu
    \end{pmatrix}
    = \begin{pmatrix}
            1 & 0 & \frac{\epsilon cw}{\sqrt{1-\epsilon^2}} \\
            0 & 1 & -\frac{\epsilon sw}{\sqrt{1-\epsilon^2}} \\
            0 & 0 & - \frac{1}{\sqrt{1-\epsilon^2}} \\
      \end{pmatrix}
      \begin{pmatrix}
        A'_\mu \\
        Z'_\mu \\
        X'_\mu
    \end{pmatrix}.
\end{equation}
% where, we assume the transformation such that, the off-diagonal diagonal terms in the new basis are 0, while all the diagonal terms are 1. Since, there are 9 such independent parameters, and 6 such conditions, so we are left with independent parameters. We choose those 3 parameters by considering that $\hat{Z}$ has no component of $A'$, and $\hat{X}$ has no component along $A'$ and $Z'$. 
In this new basis, the Lagrangian of Eq.~\eqref{eq:Lagrangian} becomes
\begin{equation}
	\begin{split}
		\mathcal{L_{Z-X}} \supset & -\frac{1}{4} {A'}_{\mu\nu}{A'}^{\mu\nu}-\frac{1}{4} {Z'}_{\mu\nu}{Z'}^{\mu\nu}-\frac{1}{4} {X'}_{\mu\nu}{X'}^{\mu\nu} \\
        & +\frac{1}{2}m_Z^2 Z'_\mu Z'^\mu + \frac{1}{2} \frac{m_X^2 +m_Z^2 \epsilon^2 ~sw^2}{1-\epsilon^2} X'_\mu X'^\mu \\
        & + m_Z^2\frac{\epsilon ~sw}{\sqrt{1-\epsilon^2}} Z'_\mu X'^\mu + \frac{1}{\sqrt{1-\epsilon^2}} j_{X}^\mu {X'}_\mu\\
        & + j_{\rm em}^\mu ({A'}_\mu - \frac{\epsilon ~cw}{\sqrt{1-\epsilon^2}}X'^\mu)\\
        & + j_{Z}^\mu ({Z'}_\mu + \frac{\epsilon ~sw}{\sqrt{1-\epsilon^2}}X'^\mu)
	\end{split}
    \label{eq:Lagrangian2}
\end{equation}
yielding a mass mixing between $Z'$ and $X'$ though $A'$ field is achieved to be massless and diagonal.
% From Eqn.~\ref{eq:Lagrangian2}, we notice that the $A'$ field is massless and diagonal, but $X'$ and $Z'$ mixes up.
% In order to identify the mass basis of the gauge boson, we make a further redefinition of the fields as follows: 
Thus performing a diagonalization via a rotation angle $\eta$:
\begin{equation}
        \begin{pmatrix}
            A'_{\mu}\\
            Z'_\mu \\
            X'_\mu
        \end{pmatrix}\\
         =  \begin{pmatrix}
            1 & 0 & 0\\
            \rm cos\eta & \rm sin\eta & 0\\
            -\rm sin\eta & \rm cos\eta & 0
            \end{pmatrix}.
         \begin{pmatrix}
            A_\mu \\
            Z_\mu\\
            X_\mu
        \end{pmatrix}
\end{equation}
gives the physical masses:
\begin{equation}
    \begin{split}
        M_X^2& =\frac{m_X^2 [m_X^2 (1+\epsilon^2)-m_Z^2(1+cw^2 \epsilon^2)]}{{m_X}^2-{m_Z}^2},\\
        M_Z^2& =\frac{m_Z^2 [m_X^2-m_Z^2(1+sw^2 \epsilon^2)]}{{m_X}^2-{m_Z}^2},
    \end{split}
\end{equation}
where the mixing angle $\eta$ is given by
\begin{equation}
    \begin{split}
        \tan(2\eta)&=2\frac{m_Z^2~sw~\epsilon \sqrt{1-\epsilon^2}}{m_Z^2[1-\epsilon^2(1+sw^2)]-m_X^2}\\
        &=2\frac{\epsilon\,sw}{1-\delta} + \mathcal{O}(\epsilon^3),
    \end{split}
\end{equation}
with $\delta=m_X^2/m_Z^2$. 

% Recall that $m_X$ is the mass term of the field $\hat{X}_\mu$ generated by the Stueckelberg mechanism and $m_Z^2=\frac{(g^2+g'^2)v^2}{4}$ is the mass of the $\hat{Z}$ boson in the SM. 
% The physical masses of the $Z$ and $X$ gauge bosons are given by the following:
Thus, the interaction of the physical gauge boson $X_\mu$ with the SM fermions is described by Eq.~\eqref{eq:Lagrangian2} and is given by:
% introduces the $X_\mu$ interaction with the SM fermions $f$ (see Appendix~\ref{app:fields}), as given by
\begin{equation}
    X_\mu j^\mu_{SM} = -X_\mu \overline{f} \gamma^\mu \{ C^f_L P_L + C^f_R P_R\} f
    \label{eq:X-interaction}
\end{equation}
where $C^f_L$ and $C^f_R$ are the left-handed and the right-handed couplings of SM fermions $f$ to $X_\mu$, and they are conveniently expressed in terms of the SM couplings as follows (see Appendix \ref{app:fields})
\begin{equation}
    \begin{split}
        C^f_L &= \left( \sin\eta + \frac{\cos\eta~\epsilon~sw}{\sqrt{1-\epsilon^2}} \right)g^f_L-\left( \frac{\cos\eta~\epsilon~cw}{\sqrt{1-\epsilon^2}} \right) eQ^f, \\
        C^f_R &=\left( \sin\eta + \frac{\cos\eta~\epsilon~sw}{\sqrt{1-\epsilon^2}} \right)g^f_R-\left( \frac{\cos\eta~\epsilon~cw}{\sqrt{1-\epsilon^2}} \right) e Q^f,
    \end{split}
\end{equation}
with 
\begin{equation}
    \begin{split}
        g_L^f &= \frac{g}{cw} \left( T^{3f}_{L} -sw^2 Q^f\right), \\
        g_R^f &= -\frac{g}{cw} \left( sw^2 Q^f \right)\,.
    \end{split}
\end{equation}
Here, $g$ is the SM $SU(2)_L$ gauge coupling and $T^3_L$ is the third component of the $SU(2)_L$ isospin. 

Given the above-mentioned interactions, there are two ways the SM particles can interact with the DS particles: 1) the kinetic mixing parameter $\epsilon$, and 2) the quartic coupling $\lambda_{h\phi}$ which quantifies the strength of $h h \rightarrow \phi^{\star}\phi$ scattering. For the case of very small $\epsilon$, the quartic coupling $\lambda_{h\phi}$ serves as the only portal (Fig.~\ref{fig:portal}) through which the dark sector can thermalize with the SM bath. As shown in Eq.~\eqref{eqn::potential}, the coupling $\lambda_{h\phi}$ plays a crucial role in determining the thermalization of $\phi$. The processes through which $\phi$ can equilibrate with the SM bath are $h \leftrightarrow \phi \phi^\star$, and $hh \leftrightarrow \phi \phi^\star$. If $m_\phi$ $>$ $M_h/2$, the decay process is forbidden and the only thermalization is through $hh \leftrightarrow \phi \phi^\star$ scattering. However, if $m_\phi$ $\leq$ $M_h/2$, both the processes can produce $\phi$ in the early Universe. In either of the cases, the production is possible and the thermalization is feasible for the scattering rate $\Gamma_{h\rightarrow\phi} \geq H$, where $H$ is the Hubble expansion rate, which sets the coupling $\lambda_{h\phi}$ to a value $\geq$ $10^{-5}$.

\subsection{Dark Matter Self Interactions}

In our setup, the dark current interaction arises from the term $\frac{1}{\sqrt{1-\epsilon^2}} j_{X}^\mu {X'}_\mu$ in Eq.~\eqref{eq:Lagrangian2}. Consequently, the relevant Lagrangian for self-interacting dark matter takes the form $(g_d \cos\eta / \sqrt{1-\epsilon^2}) X_{\mu} \overline{\chi} \gamma^{\mu} \chi$ or equivalently, $g_\chi X_{\mu} \overline{\chi} \gamma^{\mu} \chi$. This interaction generates a Yukawa potential between DM particles:
$
V(r)= \pm ({\alpha_\chi}/{r})\exp[{-M_{X} r}], \quad \alpha_\chi \equiv {g^2_\chi}/{4\pi}
$
where $+$ ($-$) corresponds to repulsive (attractive) interactions. Notably, while $\chi\overline{\chi}$ interactions are attractive, $\chi\chi$ and $\overline{\chi}\overline{\chi}$ interactions are repulsive. The DM self-scattering is conventionally characterized by the transfer cross-section:
$
    \sigma_T = \int d\Omega (1 - \cos\theta) \frac{d\sigma}{d\Omega},
$
as discussed in Refs.~\cite{Feng:2009hw,Tulin:2013teo,Tulin:2017ara}. A complete analysis requires computations beyond the perturbative limit. Depending on the DM mass ($m_{\chi} \equiv m_{\rm DM}$), mediator mass ($m_{X}$), relative DM velocity ($v_{\rm rel}$), and interaction strength ($g_{\chi}$), three distinct regimes emerge:
 {Born regime:} ${g^2_{\chi} m_{\chi}}/{4\pi M_{X}} \ll 1$, $m_{\chi} v_{\rm rel} / M_{X} \geq 1$, {Classical regime:} ${g^2_{\chi} m_{\chi}}/{4\pi M_{X}} \geq 1$, {Resonant regime:} ${g^2_{\chi} m_{\chi}}/{4\pi M_{X}} \geq 1$, $m_{\chi} v_{\rm rel} / M_{X} \leq 1$.
Further details can be found in Appendix~\ref{app::sidm_cross-section}.

\begin{figure}[h]
    \centering
    \includegraphics[width=0.45\textwidth]{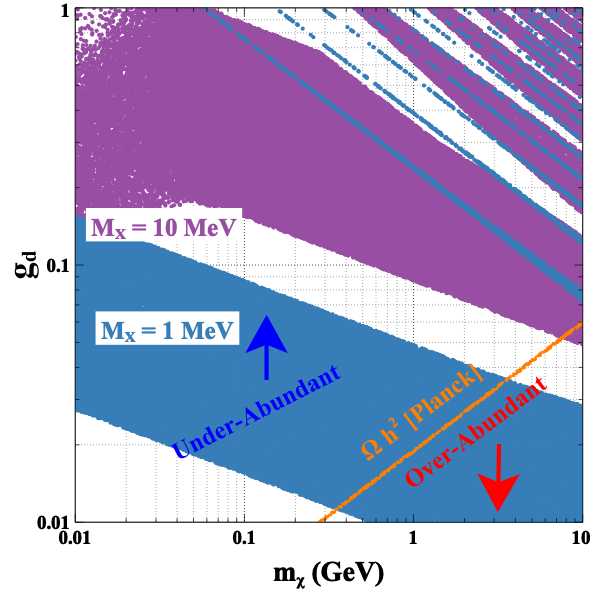}
    \caption{The SIDM parameter space in the $g_d$ vs. $m_\chi$ plane for DM self-interactions, with $\sigma/m_{\chi}$ ranging from $0.1$ to $100$ cm$^2$/g. The orange line represents $\Omega h^2 = 0.120 \pm 0.001$.}
    \label{fig:abundance}
\end{figure}

If DM reaches thermal equilibrium in the early Universe, the same Lagrangian also governs DM annihilation into $X$ mediators, contributing to its thermal relic abundance. The corresponding Feynman diagrams are given in Appendix \ref{app::th_avergae}. The dominant thermally averaged annihilation cross-section is $\langle \sigma v \rangle_{\chi \chi \to XX} \simeq \pi \alpha^2_\chi/m^2_\chi$.
% \begin{equation}
%    \langle \sigma v \rangle_{\chi \chi \to XX} = \frac{g_d^4}{16 \pi m_\chi^2} \sqrt{1 - \frac{m_X^2}{m_\chi^2}}.
%     \label{eq:dm_ann}
% \end{equation}
For all practical purposes, we set $g_\chi = g_d$, since the kinetic mixing parameters $\epsilon$ and $\eta$ are extremely small in the region of our interest. Since we focus on SIDM, the $U(1)_D$ coupling $g_d$ lies within the range $[10^{-2}, 1]$. The thermally averaged annihilation cross-section depends on $g_d$ and $m_\chi$, meaning different values of $m_\chi$ leading to varying annihilation cross-sections, which impact the relic abundance of DM.
Fig.~\ref{fig:abundance} illustrates the viable parameter space for two mediator masses, $M_X = 1~{\rm MeV}$ and $10~{\rm MeV}$, highlighting the regions where excessive DM annihilation results in a relic density below the observed value, indicated by the orange line.

\section{Dark Matter Relic}\label{sec::relic}
\subsection{Thermalization of Dark Sector particles}

The dark sector thermalizes with the visisble sector through two mechanisms: (1) the kinetic mixing parameter $\epsilon$, and (2) the quartic coupling $\lambda_{h\phi}$. When $\epsilon$ is very small, the complex scalar $\phi$ serves as the primary portal (Fig.~\ref{fig:portal}) through which the dark sector thermalizes with the SM bath. Thermalization requires $\Gamma_{H\leftrightarrow\phi} \geq H$, translating to $\lambda_{h\phi} \geq 10^{-5}$. Once thermalized, $\phi$ maintains equilibrium via $\phi\,\text{SM} \leftrightarrow \phi\,\text{SM}$ scattering until kinetic decoupling.
% There are two ways, the SM particles can interact with the DS particles: 1) the kinetic mixing parameter $\epsilon$, and 2) the quartic coupling $\lambda_{h\phi}$ which quantifies the strength of $H H \rightarrow \phi^{\star}\phi$ scattering. For the case of very small $\epsilon$, the complex scalar $\phi$ serves as the only portal (Fig.~\ref{fig:portal}) through which the dark sector can thermalize with the SM bath. As shown in Eqn.~\ref{eqn::potential}, the coupling $\lambda_{h\phi}$ plays a crucial role in determining the thermalization of $\phi$. The processes through $\phi$ equilibrates with the SM bath are $h \leftrightarrow \phi \phi^\star$, and $hh \leftrightarrow \phi \phi^\star$. If $m_\phi$ $>$ $M_h/2$, the decay process is forbidden and the only thermalization is through $hh \leftrightarrow \phi \phi^\star$ scattering. However, if $m_\phi$ $\leq$ $M_h/2$, both the processes can produce $\phi$ in the early Universe. In either of the cases, the production is possible and the thermalization is feasible for the scattering rate $\Gamma_{H\rightarrow\phi} \geq H$, where $H$ is the Hubble expansion rate, which sets the coupling $\lambda_{h\phi}$ to a value $\geq$ $10^{-5}$. Once it is produced, it can maintain a thermal equilibrium with the SM bath until it decouples kinetically via $\phi~\rm{SM} \leftrightarrow \phi~\rm{SM}$, where SM denotes the SM particles.

For the process $\phi~\rm{SM} \leftrightarrow \phi~\rm{SM}$, the interaction rate can be computed as (see Appendix~\ref{app::int_rate_scattering}) 
\begin{equation}
    \Gamma_{\rm int} = \frac{g_{\rm SM}}{4 \pi^2 m_\phi^2 K_2(\frac{m_\phi}{T})} \int_{s_{min}}^\infty \sigma(s) p(s)^2 \sqrt{s} K_1\left(\frac{\sqrt{s}}{T}\right)
\end{equation}
where $s_{\rm min}=(m_\phi + m_{\rm SM})^2)$ and the function $p(s)$ is given by
\begin{equation}
    p(s)=\frac{\left[s-(m_{\phi}+m_{\rm SM})^2\right]^{\frac{1}{2}} \left[s-(m_{\phi}-m_{\rm SM})^2\right]^{\frac{1}{2}} }{2 \sqrt{s}}.
\end{equation}
% The $\Gamma_{int}$ and $H$ are plotted as a function of temperature using $\lambda_{h\phi}$ $=$ $10^{-3}$ in Fig.~\ref{fig:Kin_Int_Rate} (top for $m_\phi=10$ GeV and bottom for $m_\phi=200$ GeV). From Fig.~\ref{fig:Kin_Int_Rate}, we see that the most relevant process which keeps $\phi$ in equilibrium until the late epoch is $b \phi \leftrightarrow b \phi$ scattering, where $b$ is the bottom quark. We also notice that the decoupling temperature is small for the smaller $m_\phi$.
\begin{figure}[h]
	\centering
    \includegraphics[width=0.45\textwidth]{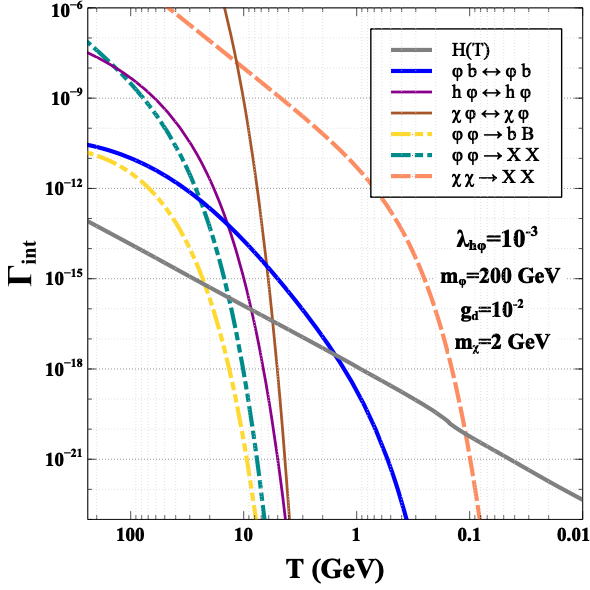}
	\caption{Kinetic decoupling of $\phi$ particle from various processes for $\phi$ mass 200 GeV. For definiteness, we take $m_\chi$=2 GeV and $g_d=0.01$.}
	\label{fig:Kin_Int_Rate}
\end{figure}

The interaction rate $\Gamma_{\rm int}$ and the Hubble parameter $H$ are plotted as functions of temperature for $\lambda_{h\phi} = 10^{-3}$, $m_\phi = 200$ GeV, $m_\chi=2$ GeV in Fig.~\ref{fig:Kin_Int_Rate}. From this plot, it is evident that the dominant process keeping $\phi$ in equilibrium with the SM at late times is $b \phi \leftrightarrow b \phi$ scattering, where $b$ represents the bottom quark. 

In addition to its quartic coupling to the SM Higgs, $\phi$ also directly couples to the $U(1)_D$ gauge boson $X_\mu$ through an interaction of the form $g_d^2 |\phi|^2 X_\mu X^\mu$. This strong coupling ensures sufficient interactions among $\phi$, $X_\mu$, and $\chi$, leading to their thermalization within the dark sector.
% Not only the $\phi$ particles couple directly to SM Higgs via a quartic coupling $\lambda_{h\phi}$, there is also a direct coupling of $\phi$ with the $U(1)_D$ gauge boson $X_\mu$ of the form given by $g_d^2 |\phi|^2 X_\mu X^\mu$. Such a large coupling can provide sufficient interactions among $\phi$, $X_\mu$ and $\chi$ and lead to thermalization among themselves.

\subsection{Chemical Decoupling of the Hidden Sector Particles}
For $\lambda_{h \phi} \geq 10^{-5}$, the scalar $\phi$ remains in thermal equilibrium with the SM bath, which in turn thermalizes dark matter $\chi$. From Fig.~\ref{fig:abundance}, we observe that for sub-GeV DM masses, thermal relics are underabundant across most of the parameter space where self-interactions are significant ($g_d \geq 0.01$). This deficiency is compensated by the late decay of $\phi$ after its freeze-out. Here it is worth noticing that, the freeze-out abundance of $\phi$ is also controlled by the the same coupling $g_d$ along with its mass. The $\phi$ and $\chi$ number density changing processes are given in Appendix \ref{app::th_avergae}, through which they decouple from the thermal bath and produce their thermal relic abundances. The thermally averaged annihilation cross-section for $\phi \phi \leftrightarrow XX$ is $ \left< \sigma v\right>_{\phi \phi \rightarrow X X}\simeq{6 \pi \alpha_d}/{m_\phi^2} $(see Appendix~\ref{app::th_avergae} for details). 

\begin{figure}[h]
	\centering
	\includegraphics[width=0.45\textwidth]{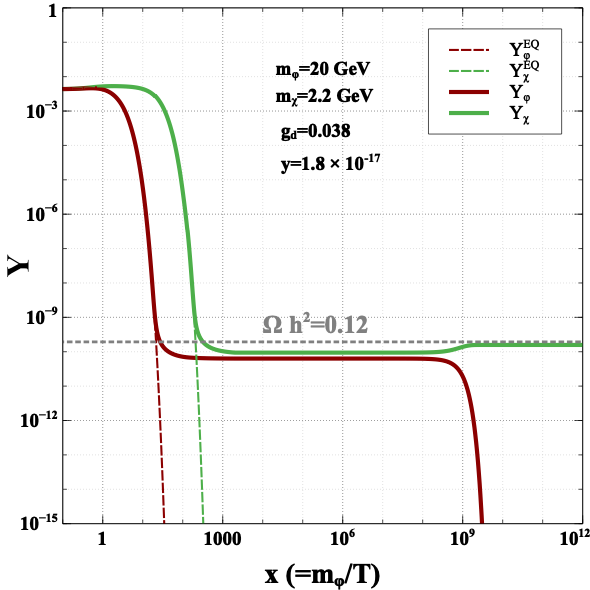}
	\caption{The evolution of the co-moving number densities of $\phi$ and $\chi$ as a function of $x (=m_\phi/T)$.}
	\label{fig:Relic_Plot}
\end{figure}

To track the evolution of the co-moving number densities of $\phi$ and $\chi$, we solve the coupled Boltzmann equations considering all the number changing processes into account.  The details are given in Appendix~\ref{app::complete_BE}. A representative solution for a benchmark set of parameters is shown in Fig.~\ref{fig:Relic_Plot}, where the solid red and green curves correspond to the co-moving number densities of $\phi$ and $\chi$, respectively. The dashed lines depict their respective equilibrium abundances, while the gray dotted horizontal line represents the observed relic abundance of $\chi$ for $m_\chi = 2.2$ GeV. The relevant parameter choices are indicated in the plot. As expected, for $m_\chi = 2.2$ GeV and $g_d = 0.038$, the thermal component of $\chi$ is underabundant. This deficit is compensated by the late-time decay of $\phi$.

It is important to note that the decoupling and freeze-out abundances of both $\phi$ and $\chi$ are primarily governed by the parameters $g_d$, $m_\phi$, and $m_\chi$, with the $\phi$ abundance depending on $g_d$ and $m_\phi$, and the $\chi$ abundance controlled by $g_d$ and $m_\chi$. On the other hand, the late-time decay of $\phi$, which converts its thermal relic into a non-thermal component of DM $\chi$, is dictated by the parameters $y$ and $m_\phi$. Therefore, achieving the correct total DM relic abundance for a fixed $g_d$ and $m_\chi$ requires a careful adjustment of $m_\phi$ to ensure that the non-thermal contribution from $\phi$ decay does not result in an overabundance of $\chi$ at late times.

To identify the viable parameter space that yields the correct DM relic density, we implement a scanning procedure with the following strategy. For fixed values of $g_d$ and $M_X$, we first determine the range of $m_\chi$ for which a purely thermal relic is insufficient to satisfy the observed relic density, while still respecting the self-interaction requirements for SIDM. For each such $m_\chi$, we compute the thermal relic abundance of $\chi$ and quantify the resulting shortfall. With this information, we search for the corresponding $m_\phi$ that produces the appropriate thermal relic of $\phi$ such that its subsequent decay into $\chi$ and $\nu_S$ fills the deficit, thereby fulfilling the observed DM relic criteria. It is also worth mentioning that since $M_X$ is chosen to be of the order $\mathcal{O}(1)$ MeV, it does not significantly impact the freeze-out abundances of $\phi$ and $\chi$, which lie in the $\mathcal{O}(10)$ GeV and $\mathcal{O}(1)$ GeV regimes, respectively.

% The decoupling and subsequent decay of $\phi$ depends on both $g_d$ and $m_\phi$, where the decay provides a non-thermal contribution to the relic density of $\chi$. Care must be taken to ensure that the decay does not lead to an overproduction of $\chi$. As seen in Fig.~\ref{fig:freezeINfraction}, for a fixed $m_\phi$ (e.g., 30 GeV), different choices of $g_d$ lead to varying freeze-out abundances of $\phi$: approximately 30\% for $g_d = 0.04$ and 50\% for $g_d = 0.03$. This abundance directly sets the non-thermal relic yield of $\chi$ via $\phi$ decay.

% The remaining contribution to the observed DM relic must then come from the thermal component, which is governed by the annihilation cross-section (see Eq.~\ref{eq:dm_annihilation}), and hence depends on $g_d$ and $m_\chi$. Since $g_d$ is already fixed in this analysis, a lower $g_d$ (which corresponds to a higher $\phi$ relic) requires a smaller thermal relic from $\chi$. This can be achieved if $\chi$ has a stronger annihilation rate, which naturally happens for lighter $\chi$ masses. Therefore, for smaller $g_d$, the viable range of $m_\chi$ shifts to lower values. As a result, the non-thermal fraction of $\chi$ can vary between 1\% to nearly 100\%, depending on the parameter choices.

\begin{figure}[h]
	\centering
	\includegraphics[width=0.45\textwidth]{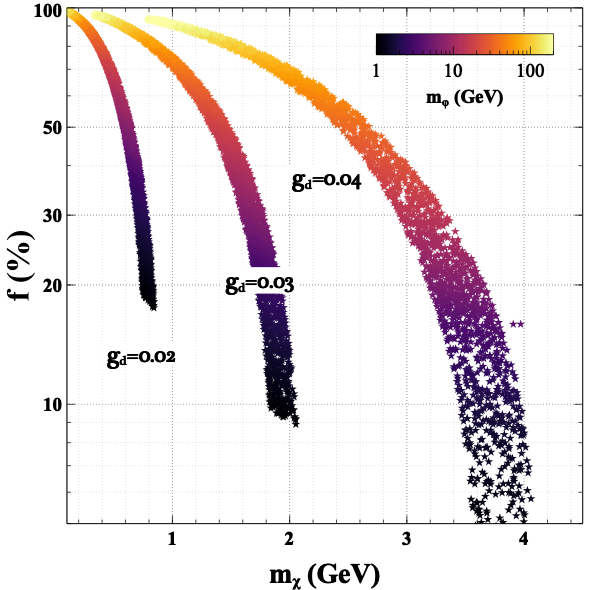}
	\caption{The fraction of DM relic, $f={Y_\chi^{\rm nth}}/{(Y_\chi^{\rm th}+Y_\chi^{\rm nth})}$, shown as a function of $m_\chi$. Here, ${\rm th}$ and ${\rm nth}$ in superscripts denote thermal and non-thermal components, respectively, under the constraint $(Y_\chi^{\rm th}+Y_\chi^{\rm nth}) = Y_{\chi}^{\rm Obs}$.}
	\label{fig:freezeINfraction}
\end{figure}

The results of our parameter scan are presented in Fig.~\ref{fig:freezeINfraction}, where we show the non-thermal dark matter relic fraction $f = {Y_\chi^{\rm nth}}/{Y_\chi^{\rm Obs.}}$ as a function of the dark matter mass $m_\chi$ for three different values of $g_d$. The color gradient in the plot represents the corresponding values of $m_\phi$ required to obtain the correct relic density. 
We observe that, for a fixed $g_d$, increasing $m_\chi$ leads to a gradual decrease in the non-thermal fraction $f$, accompanied by a corresponding decrease in $m_\phi$. This behavior can be understood as follows: as $m_\chi$ increases, the annihilation cross-section $\langle \sigma v \rangle \propto 1/m_\chi^2$ decreases, resulting in a larger thermal contribution to the DM relic abundance. Consequently, a smaller non-thermal contribution is needed to match the observed relic density. 
To compensate, the thermal abundance of $\phi$, which decays into $\chi$ must also be reduced. Since the annihilation cross-section of $\phi$  scales as $\langle \sigma v \rangle \propto 1/m_\phi^2$, achieving a smaller $\phi$ relic requires a lower $m_\phi$, explaining the observed trend.

\section{$\boldsymbol{N}_{\rm \bf eff}$ from $\boldsymbol{\phi}$ decay}\label{sec::Neff}

\begin{figure}[h]
	\centering
	\includegraphics[width=0.25\textwidth]{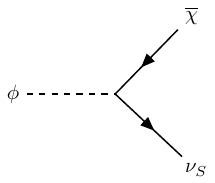}
	\caption{Feynman diagram of the decay $\phi$ to $\chi$ and $\nu_S$.}
	\label{fig:phi_decay}
\end{figure}

The late time decay of the scalar $\phi$ produces both the dark matter candidate $\chi$ and the dark radiation component $\nu_S$, as illustrated in Fig.\ref{fig:phi_decay}. This process contributes non-thermally to the effective number of relativistic species, $N_{\rm eff}$. We are specifically interested in the scenario where the $\phi$ decay occurs after the epoch of BBN but before the epoch of CMB formation. The decay width of $\phi$ is given by:
\begin{equation}
\Gamma_\phi = \frac{y^2 m_\phi}{16 \pi} \left( 1-\frac{m_\chi^2}{m_\phi^2} \right)^2
\label{eq:phi_decay_width}
\end{equation}

After $\phi$ freezes out and becomes non-relativistic, its energy density $\rho_\phi$ scales as $a^{-3}$. Therefore, relative to the background radiation density $\rho_{\rm rad}$ (which scales as $a^{-4}$), the energy density of $\phi$ effectively increases over time. Although a long-lived $\phi$ could, in principle, come to dominate the energy density of the Universe, in our scenario it decays such that $\rho_\phi$ never overtakes radiation before its decay. Consequently, if $\phi$ decays when the ratio $\rho_\phi/\rho_{\rm rad}$ is small, the injected dark radiation (and hence the change in $N_{\rm eff}$) will be minimal. Conversely, a later decay occurring when $\rho_\phi/\rho_{\rm rad}$ is relatively larger, yields a higher contribution to $N_{\rm eff}$.
Thus if $\phi$ decays shortly after BBN, the contribution of $\nu_S$ to $N_{\rm eff}$ remains small. However, if the decay occurs near the CMB epoch, the contribution can be significantly enhanced and potentially detectable in future CMB experiments. However, relativistic $\chi$ particles produced non-thermally near the CMB epoch may disrupt structure formation~\cite{Boehm:2004th}. Thus the Yukawa coupling $y$ must be carefully chosen to ensure that decay of $\phi$ satisfies the $\chi$ relic density, provides an observable change in $N_{\rm eff}$, and complies with structure formation constraints. 

The additional relativistic degrees of freedom $\Delta N_{\rm eff}$ at the time of CMB, can be expressed as
\begin{equation}
    \Delta N_{\rm eff}=N_{\rm eff}-N^{\rm SM}_{\rm eff} = N_{\nu_S}\times \left. \frac{\rho_{\nu S}}{\rho_{\nu L}} \right|_{T_{\rm CMB}}
    \label{eq:neff_non-th}
\end{equation}
where $\rho_{\nu_L}=2~ \frac{7}{8} \frac{\pi^2}{30} T_{\nu{_L}}^4$ denotes the energy density of one generation of light neutrinos ($\nu_L$) is the energy density of one generation of SM neutrinos, and $\rho_{\nu_S}$ denotes the corresponding energy density of the sterile neutrino. $N_{\nu_S}$ is the number of $\nu_S$ generations. If neutrinos decouple instantaneously, then $N^{\rm SM}_{\rm eff}=3$. Considering non-instantaneous decoupling of SM neutrinos along with flavour oscillations and plasma correction of quantum electrodynamics, the $N^{\rm SM}_{\rm eff}$ value shifts to $N_{\rm eff}^{\rm SM} = 3.045$ \cite{Mangano:2005cc, Grohs:2015tfy,deSalas:2016ztq}\footnote{A few recent works \cite{Froustey:2020mcq, Bennett:2020zkv, Drewes:2024wbw} obtained a slightly different prediction $N_{\rm eff}^{\rm SM} = 3.044$.}. To track the evolution of the energy density of $\nu_S$, we solve the Boltzmann equation:
\begin{equation}
    \frac{d\rho_{\nu_S}}{dx}=-\frac{4 \beta \rho_{\nu_S}}{x} + \frac{1}{x H(x)} \langle E \Gamma_{\phi} \rangle (Y_{\phi}~s(x))
    \label{eq:radiation}
\end{equation}
where {$\beta = 1+T\frac{dg_s/dT}{3 g_s}$} and $g_s$ is the effective relativistic degrees of freedom related to the entropy density $s$. The thermal-averaged energy transfer rate $\langle E \Gamma_{\phi} \rangle$ is given by \cite{Biswas:2022vkq}:
\begin{equation}
\langle E \Gamma_{\phi} \rangle =g_\chi g_{\nu_S}\frac{|\mathcal{M}|_{\phi \rightarrow \chi \nu_S}^2}{32 \pi m^2{\phi}} (m_\phi^2-m_\chi^2) \left( 1-\frac{m_\chi^2}{m_\phi^2} \right).
\end{equation}

\begin{figure}[h]
	\centering
	\includegraphics[width=0.45\textwidth]{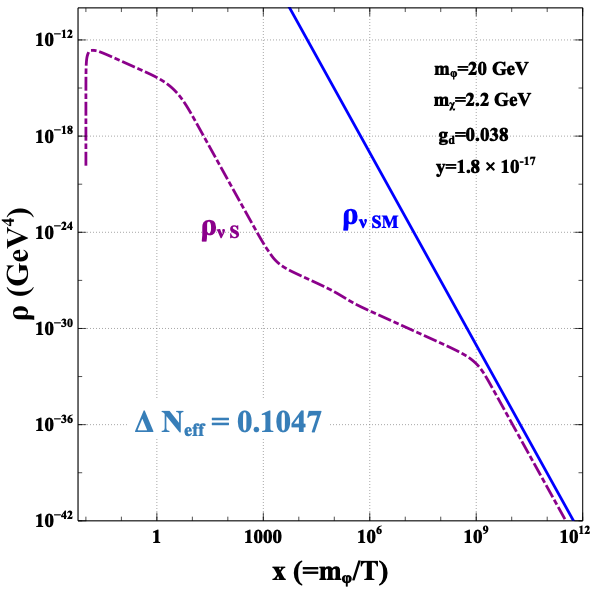}
	\caption{The evolution of the energy density of $\nu_S$ and SM neutrinos illustrate their significant contribution to $N_{\rm eff}$ at the time of recombination epoch.}
	\label{fig:Radiation_Plot}
\end{figure}

We numerically solve Eq. \eqref{eq:radiation} together with the Boltzmann equations for DS particles $X, \chi, \phi$ given in Appendix \ref{app::complete_BE} simultaneously to track the evolution of $\rho_{\nu_S}$ along with the comoving abundances of $\phi$ and $\chi$. Using the same benchmark parameter set employed in Fig. \ref{fig:Relic_Plot} for the $\phi$ and $\chi$ evolution, the resulting evolution of $\rho_{\nu_S}$ is shown in Fig. \ref{fig:Radiation_Plot}. The corresponding contribution to $\Delta N_{\rm eff}$ is then computed using Eq.~\eqref{eq:neff_non-th} and found to be 0.1047.

\begin{figure}[h]
	\centering
    \includegraphics[width=0.45\textwidth]{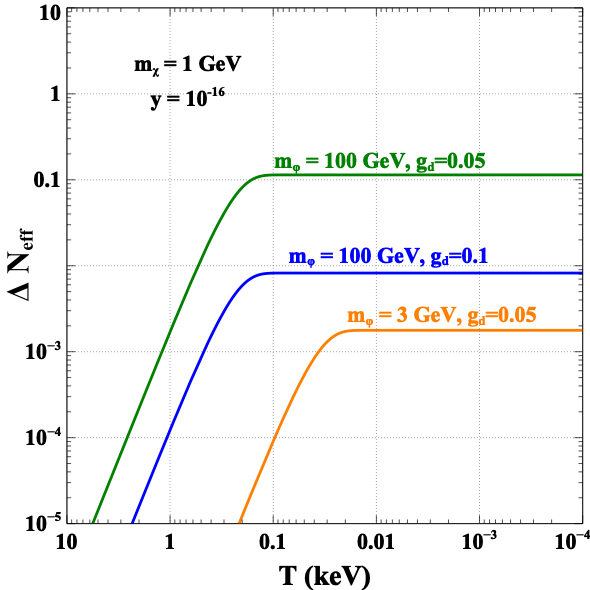}
	\caption{Effect on $\Delta N_{\rm eff}$ from the key parameters $g_d$ and $m_{\phi}$. Here, we fixed $m_\chi=1$ GeV and $y=10^{-16}$.}
	\label{fig:Neff}
\end{figure}

To explore the sensitivity of $\Delta N_{\rm eff}$ to the model parameters $g_d$, $m_\phi$, and $y$, we perform a random scan, ensuring in each case that the correct relic abundance for $\chi$ is reproduced. The variation of $\Delta N_{\rm eff}$ for three representative combinations of ${m_\phi, g_d}$ is illustrated in Fig.~\ref{fig:Neff}, while fixing $m_\chi = 1$ GeV and $y = 10^{-16}$.
The plot clearly demonstrates two key trends: (i) a smaller $g_d$ leads to a larger thermal relic abundance of $\phi$, which enhances the subsequent $\Delta N_{\rm eff}$, and (ii) increasing $m_\phi$ raises the energy density of $\phi$ ({\it i.e.} $\rho_\phi = m_\phi n_\phi$) (in addition to increasing freeze-out abundance of $\phi$ number density), which thus amplifies the energy density of DR $\nu_S$ when $\phi$ decays to $\nu_S$ thereby enhancing $\Delta N_{\rm eff}$. This can be seen by comparing the orange and green curves, both corresponding to $g_d = 0.05$, where increasing $m_\phi$ from 3 GeV (orange) to 100 GeV (green) significantly boosts $\Delta N_{\rm eff}$. Likewise, fixing $m_\phi = 100$ GeV and comparing the green ($g_d = 0.05$) and blue ($g_d = 0.1$) curves shows that reducing $g_d$ enhances $\Delta N_{\rm eff}$, as expected.

\begin{figure}[h]
	\centering
    \includegraphics[width=0.45\textwidth]{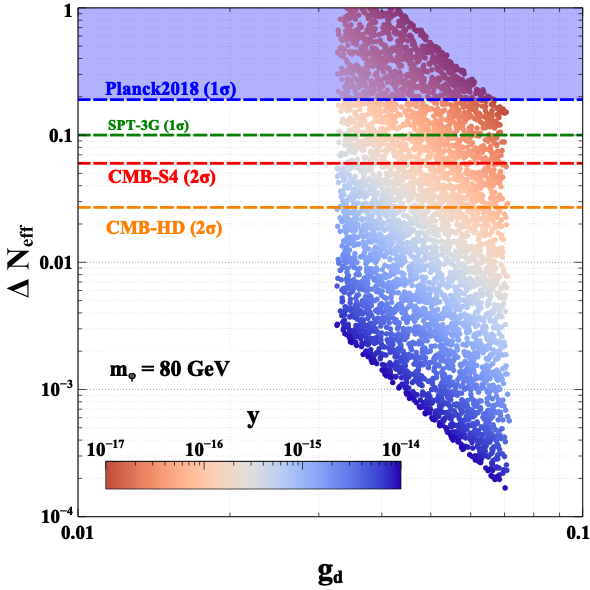}
	\caption{Contribution to $\Delta N_{\rm eff}$ as a function of $g_d$ for $m_\phi=80$ GeV. The Yukawa coupling $y$ is varied as shown in the color bar.}
	\label{fig:Neffvsgd}
\end{figure}
\begin{figure}[h]
	\centering
    \includegraphics[width=0.45\textwidth]{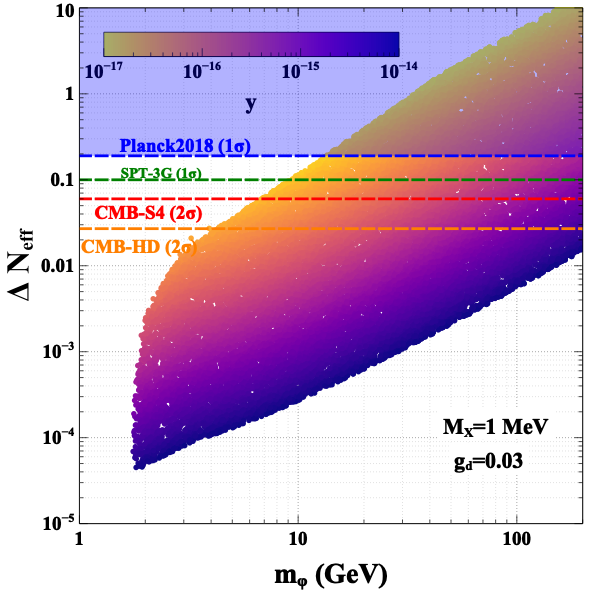}
	\caption{The parameter space that will give the $\Delta N_{\rm eff}$ from the process $\phi \rightarrow \chi \nu_S$ and the correct relic of the dark matter. All the points shown above satisfy the correct DM relic ballpark for $m_\chi$, as shown in Fig.~\ref{fig:freezeINfraction}. Note that different values of $g_d$ allows different ranges of Yukawa coupling $y$ for the oservable $\Delta N_{\rm eff}$.}
	\label{fig:Neff_vs_mphi}
\end{figure}

In Fig. \ref{fig:Neff}, the Yukawa coupling is fixed at $y = 10^{-16}$, while different values of $m_\phi$ and $g_d$ are chosen to study their impact on $\Delta N_{\rm eff}$. 
To further illustrate this dependence as well as to see the effect of the Yukawa coupling, Fig.\ref{fig:Neffvsgd} presents the variation of $\Delta N_{\rm eff}$ as a function of $g_d$ for a fixed $m_\phi = 80$ GeV, while Fig.~\ref{fig:Neff_vs_mphi} shows $\Delta N_{\rm eff}$ as a function of $m_\phi$ for a fixed $g_d = 0.03$ with the Yukawa coupling $y$ varied as indicated by the color gradient in both the figures.
As explained before, for a fixed $y$, $\Delta N_{\rm eff}$ decreases with increasing $g_d$ (Fig.\ref{fig:Neffvsgd}) and increases with increasing $m_\phi$ (Fig.\ref{fig:Neff_vs_mphi}). Furthermore, for a fixed $g_d$ or $m_\phi$, a smaller Yukawa coupling $y$ results in a larger $\Delta N_{\rm eff}$. This is because a smaller $y$ implies a longer $\phi$ lifetime, allowing its energy density to grow relative to the background radiation before decay, thereby contributing more significantly to $\Delta N_{\rm eff}$.
Throughout this parameter scan, the dark matter mass $m_\chi$ has been varied within the range of [1–10] GeV for both Fig. \ref{fig:Neffvsgd} and Fig. \ref{fig:Neff_vs_mphi}. For reference, both the figures also display the $1\sigma$ upper bound on $\Delta N_{\rm eff}$ from Planck~\cite{Planck:2018vyg} and the projected sensitivities of upcoming CMB experiments such as SPT-3G~\cite{SPT-3G:2019sok} (green line), CMB-S4~\cite{Abazajian:2019eic} (red line), and CMB-HD~\cite{CMB-HD:2022bsz} (orange line).
Here, it is worth noting that we restrict the mass of the mediator $M_X$ to be $1$ MeV or below, ensuring that it can decay only into SM neutrinos. This choice allows the scenario to safely evade stringent bounds from indirect detection experiments, as well as the constraints from CMB measurements on dark matter annihilation into charged final states mediated by $X$ bosons~\cite{Madhavacheril:2013cna,Slatyer:2015jla,Planck:2018vyg, Elor:2015bho, Profumo:2017obk} of the type $\chi \bar{\chi} \to X X \to 4~{\rm SM}$.

% In Fig.~\ref{fig:Neff}, the Yukawa coupling $y$ was fixed at $y=10^{-16}$, while $m_\phi$ and $g_d$ are varied. Now, we fix one of the key parameter $m_\phi=80$ GeV and calculate $\Delta N_{\rm eff}$ as a function of $g_d$ in Fig.~\ref{fig:Neffvsgd}. On the other hand, in Fig.~\ref{fig:Neff_vs_mphi}, we fix the other key parameter $g_d=0.03$, and calculated $\Delta N_{\rm eff}$ as a function of $m_\phi$. In both the figures, the Yukawa coupling $y$ is varying in the color code. As expected, we see that $\Delta N_{\rm eff}$ decreases as $g_d$ increases from left to right as shown in Fig.~\ref{fig:Neffvsgd}. Similarly, we also observe that $\Delta N_{\rm eff}$ increases as $m_\phi$ increases from left to right as shown in Fig.~\ref{fig:Neff_vs_mphi}. Then, for a given $g_d$ (Fig.~\ref{fig:Neffvsgd}) or a given $m_\phi$ (Fig.~\ref{fig:Neff_vs_mphi}), a smaller Yukawa coupling $y$ gives rise to a larger $\Delta N_{\rm eff}$ and vice-versa. It is also to be noted that while performing the scan, we keep the DM mass in the range [1-10] GeV in both Fig.~\ref{fig:Neffvsgd} and Fig.~\ref{fig:Neff_vs_mphi}. In both the figures, we also showcase the 1 $\sigma$ upper bound from Planck's observation \cite{Planck:2018vyg}, while the green, the red and the orange lines correspond to sensitivity projection of upcoming surveys from SPT-3G \cite{SPT-3G:2019sok}, CMB-S4 \cite{Abazajian:2019eic} and CMB-HD \cite{CMB-HD:2022bsz}.

\section{Dark Matter Direct Detection}\label{sec::dd}
\begin{figure}[h]
	\centering
	\includegraphics[width=0.3\textwidth]{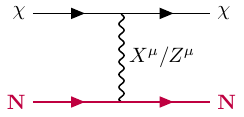}
	\caption{Feynman Diagram for the realization of dark matter detection in terrestrial laboratory. Here, $N$ denotes the nucleons.}
	\label{fig:dm_detection}
\end{figure}

In this framework, dark matter can scatter off nucleons and electrons in terrestrial detectors either through gauge boson-mediated interactions induced by kinetic mixing, or via scalar portal interactions generated at the loop level with $\nu_S$ and $\phi$ circulating in the loop. However, the loop-induced scattering cross-section remains extremely suppressed due to both the smallness of the Yukawa coupling $y$ and the loop factor. Therefore, in this work, we focus solely on the gauge portal interaction. The corresponding Feynman diagram is shown in Fig.~\ref{fig:dm_detection}. 

The effective interaction governing spin-independent DM-nucleon scattering via gauge boson mediation is described by the Lagrangian~\cite{Chun:2010ve}:
\begin{equation}
    \mathcal{L}_{\rm eff}=b_q (\overline{\chi}\gamma_\mu \chi) (\overline{q}\gamma^\mu q),
\end{equation}
where the coupling $b_q$ is defined as
\begin{equation}
    b_q = \frac{g_d (C^{qX}_L + C^{qX}_R)}{2 M_X^2}.
\end{equation}
At the nucleon level, the effective interactions with protons and neutrons can be written as:
\begin{equation*}
    \mathcal{L}_p = b_p (\overline{\chi} \gamma_\mu \chi)(\overline{p} \gamma^\mu p), \quad 
    \mathcal{L}_n = b_n (\overline{\chi} \gamma_\mu \chi)(\overline{n} \gamma^\mu n),
\end{equation*}
where the effective couplings $b_p$ and $b_n$ are related to the quark-level couplings by:
\begin{equation*}
    b_p = 2b_u + b_d, \quad b_n = b_u + 2b_d.
\end{equation*}
With these expressions, the spin-independent DM-nucleon elastic scattering cross-section is given by:
\begin{equation}
    \sigma_{\rm SI} = \frac{1}{64 \pi} \frac{\mu_r^2}{A^2} \left[ Z b_p + (A - Z) b_n \right]^2,
\end{equation}
where $\mu_r$ is the reduced mass of the DM-nucleus system, $A$ is the atomic mass number, and $Z$ is the atomic number of the target nucleus.

\begin{figure}[h]
	\centering
	\includegraphics[width=0.45\textwidth]{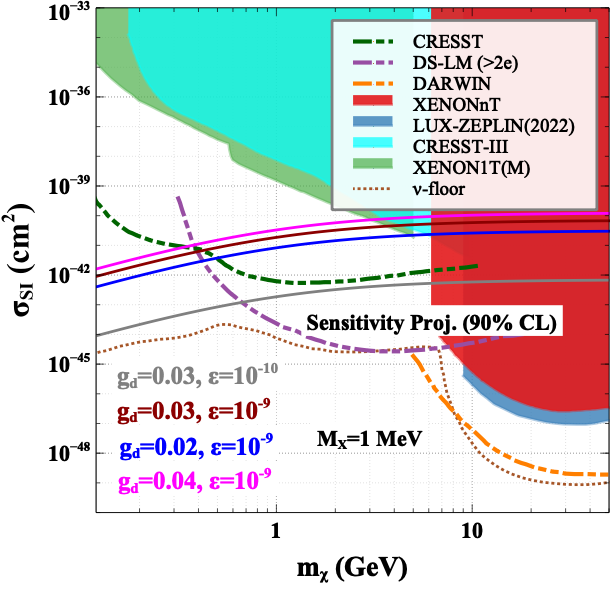}
	\caption{Spin-independent DM-nucleon cross-section for the direct detection is shown as a function of the DM mass, for a fixed $M_X$=1 MeV.}
	\label{fig:dm_detectionrate}
\end{figure}

We present the spin-independent DM-nucleon scattering cross-section as a function of dark matter mass in Fig.~\ref{fig:dm_detectionrate}, for various choices of the gauge coupling $g_d$ and kinetic mixing parameter $\epsilon$. The plot also includes existing exclusion limits from XENON1T(M)~\cite{XENON:2019zpr} (green shaded), XENONnT~\cite{XENON:2023cxc} (red shaded), CRESST~\cite{CRESST:2019jnq} (cyan region), and LUX-ZEPLIN~\cite{LZ:2022lsv} (light blue region). 
Additionally, future experimental sensitivities are illustrated using dashed lines: DSLM~\cite{GlobalArgonDarkMatter:2022ppc} (purple), CRESST~\cite{Billard:2021uyg} (dark green), and DARWIN~\cite{DARWIN:2016hyl,Schumann:2015cpa} (orange).
In this analysis, we consider $\epsilon = 10^{-9}$ or smaller, which remains consistent with constraints on the dark-photon search experiments as well as from  SN1987A constraints~\cite{Fabbrichesi:2020wbt, Chang:2016ntp} for the chosen mediator mass $M_{X}=1$ MeV. Since the direct detection cross-section scales with the gauge coupling $g_d$, lower values of $g_d$ allow the model to evade current experimental bounds particularly for dark matter masses below $3$ GeV.
However, for $m_\chi \gtrsim 3~\text{GeV}$, the parameter set with $\epsilon = 10^{-9}$ is already excluded by current direct detection experiments. A reduction of the kinetic mixing parameter to $\epsilon = 10^{-10}$ relaxes this constraint, allowing dark matter masses up to around $6$ GeV to remain viable, which is represented by the gray curve in Fig.~\ref{fig:dm_detectionrate}.

\begin{figure}[h]
	\centering
	\includegraphics[width=0.45\textwidth]{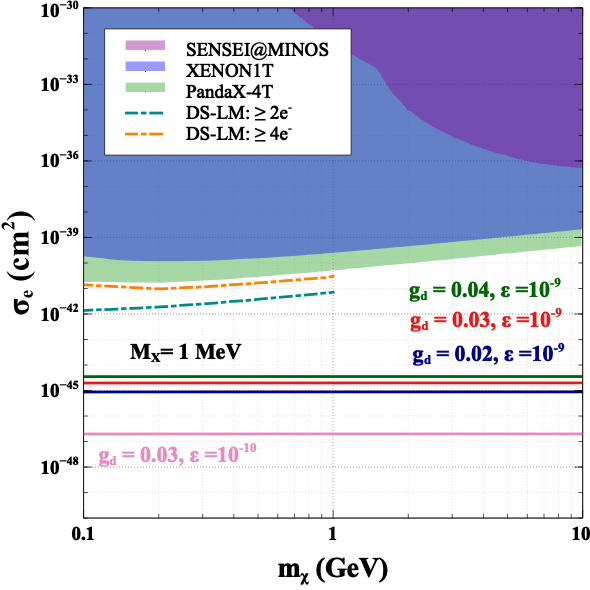}
	\caption{DM-electron cross-section for the direct detection is shown as a function of the DM mass, for a fixed $M_X$=1 MeV.}
	\label{fig:dm_e_detection}
\end{figure}

The recoil energy deposited by DM during scattering off a target particle depends on the DM mass. For light DM, scattering off electrons becomes more favorable than off heavier nuclei due to more efficient energy transfer. The cross-section for DM–electron scattering ($\chi e \leftrightarrow \chi e$) in our model is given by \cite{Hochberg:2014dra}
\begin{equation}
    \sigma_{\chi e \leftrightarrow \chi e}=\frac{\mu_{\chi e}^2}{\pi M_X^4}~g_d^2~g^2_{_{Ve}}~|F(q)|^2
\end{equation} 
where the vector coupling is defined as $g_{_{Ve}} = (C^e_L + C^e_R)/2$, as shown in Eq.~\eqref{eq:X-interaction}, $\mu_{\chi e}$ is the DM–electron reduced mass, and $|F(q)| = 1$ since $M_X \gg \alpha m_e$.  

Fig.~\ref{fig:dm_e_detection} displays existing experimental bounds from XENON1T \cite{XENON:2019gfn} (blue shaded), PandaX-4T \cite{PandaX:2022xqx} (green shaded), and SENSEI@MINOS \cite{SENSEI:2020dpa} (dark-magenta shaded). Projected sensitivity (90\% CL) from DSLM \cite{GlobalArgonDarkMatter:2022ppc} is also displayed for $4e^-$ threshold (deorange dashed line) and $2e^-$ threshold (dark cyan dashed line). Our model predictions are shown as dark green, red, and dark blue lines corresponding to $\epsilon = 10^{-9}$ and $g_d = 0.04$, $0.03$, and $0.02$, respectively. The light pink line represents the case with $g_d = 0.03$ and $\epsilon = 10^{-10}$.
Since the DM mass is much larger than the electron mass, the reduced mass $\mu_{\chi e} \approx m_e$, leading to a direct detection cross-section that remains constant across the DM mass range, as reflected in the plot.

\section{Structure Formation Constraint}\label{sec::additional}
In our framework, dark matter is produced non-thermally from the decay of a scalar field $\phi$ occurring after BBN but prior to the CMB epoch. Assuming the distribution of $\phi$ to be a delta function \cite{Aoyama:2014tga} justified at an epoch long after its thermal freeze-out, the maximum physical momentum imparted to $\chi$ is given by:
\begin{equation}
    p^{\rm max}_{\rm phys}=\frac{m_\phi^2 - m_\chi^2}{2 m_\phi}.
\end{equation}
This corresponds to a comoving momentum $p = a(t) \, p_{\text{phys}}^{\text{max}}$. The velocity of the daughter particle at any time $t$ is then given by:
\begin{equation}
    v_\chi(t) = \frac{p_{\text{phys}}(t)}{E(t)}, \quad E(t) = \sqrt{p^2_{\text{phys}}(t) + m_{\chi}^2}\,.
\end{equation}
% The DM $\chi$, with velocity $v(t)$, travels a distance known as the free-streaming length (FSL, $\lambda_{\rm fs}$) between the conformal times \(\tau = 0\) and \(\tau\), as expressed by

The comoving distance that the DM $\chi$ travels between its production time and conformal time $\tau$  is calculated by integrating the velocity as~\cite{Long:2024imw}:
\begin{equation}
    \begin{split}
        \boldsymbol{\lambda}_{\rm fs} & = \int_{\tau_{\rm dec}}^{\tau} d\tau \boldsymbol{v}_\chi(t(\tau'))\\
        & = \int_{a_{\rm dec}}^{a} \frac{da'}{a'} \frac{1}{a' H(a')} \frac{\boldsymbol{p}}{\sqrt{\boldsymbol{p}^2+a'^2 m_{\chi}^2}}\\
        & = \int_{a_{\rm dec}}^{a} \frac{da'}{a'} \frac{1}{a' H(a')} \frac{1}{\sqrt{1+(a'/a_{\rm nr})^2}}
    \end{split}
\end{equation}
Here, $t_{\rm dec}$ correspond to the time of $\phi$ decay, with $a_{\rm dec}$ as its associated scale factor. The quantity $a_{\rm nr} = p/m_{\chi}$ denotes the scale factor at which the dark matter particle $\chi$ transitions from being relativistic to non-relativistic, and $H(a)$ represents the Hubble expansion rate as a function of scale factor.

\begin{figure}[h]
	\centering
	\includegraphics[width=0.45\textwidth]{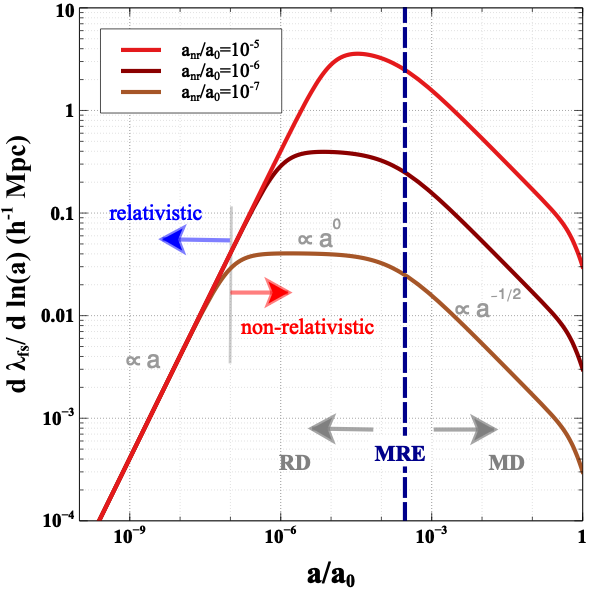}
	\caption{Contributions to the free-streaming length integral.}
	\label{fig:fsl_regions}
\end{figure}

Fig.~\ref{fig:fsl_regions} illustrates the evolution of $d(\lambda_{\rm fs})/d\ln(a) = v_\chi/(a H(a))$ as a function of the scale factor $a$ for different values of $a_{\text{nr}}$, which marks the transition of the dark matter particle $\chi$ from relativistic to non-relativistic.
While the particle remains relativistic ($a \leq a_{\text{nr}}$), its velocity $v_\chi \approx 1$ and the Hubble rate scales as $H \propto a^{-2}$ during the radiation-dominated era, making the integrand scale linearly with $a$. As the Universe expands and the particle becomes non-relativistic ($a_{\text{nr}} < a < a_{\rm eq}$), the velocity redshifts as $v_\chi \propto a^{-1}$, while $H$ still follows the radiation-dominated scaling $H \propto a^{-2}$, causing the integrand to remain approximately constant ($a^0$). 
Beyond the epoch of matter-radiation equality ($a > a_{\rm eq}$), the Universe enters the matter-dominated era, where $H \propto a^{-3/2}$ while the DM particle continues to redshift as $v_\chi \propto a^{-1}$. This results in the integrand scaling as $a^{-1/2}$. 
It is worth noting that the majority of the free-streaming length is accumulated during the period when the particle is non-relativistic but the Universe is still radiation dominated, i.e., in the range $a_{\text{nr}} < a < a_{\rm eq}$.

In a generic two-body decay, the velocity imparted to the daughter particle at production known as the kick velocity is given by: 
\begin{equation}
  v_{\rm kick}=  |v_{\chi}|=\frac{m_{\phi}^2-m_{\chi}^2}{m_{\phi}^2+m_{\chi}^2}
\end{equation}
The kick velocity, dependent on $m_\phi$ and $m_\chi$, can vary from relativistic to non-relativistic values depending on the specific mass hierarchy. Whether this velocity leaves an imprint on cosmological structure formation is determined by the associated free-streaming length, which is constrained by Lyman-$\alpha$ observations and small-scale structure data.
These constraints require: $\lambda_{\rm fs}<0.1$ Mpc to ensure consistency with the observation~\cite{DES:2020fxi, Villasenor:2022aiy,Irsic:2017ixq}.

\begin{figure}[h]
	\centering
    \includegraphics[width=0.45\textwidth]{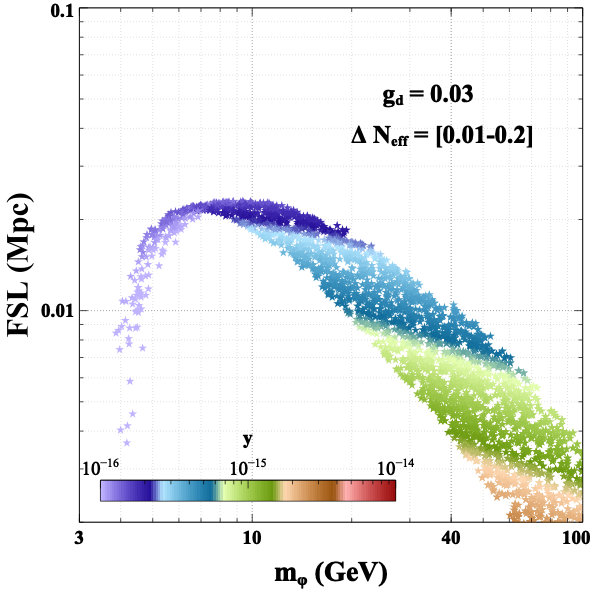}
	\caption{Model parameters that gives $\lambda_{\rm fs}<0.1$ Mpc, while giving the correct dark matter relic density and the observable $N_{\rm eff}$.}
	\label{fig:FSL}
\end{figure}

As $a_d \ll a_{\text{nr}} < a_{\text{eq}}$,  where $a_{\text{eq}}$ is the scale factor at {matter-radiation equality}, the free-streaming length is given by \cite{Kolb:1990vq, Lin:2000qq,Choi:2023jxw}:
\begin{equation}
    \begin{split}
        \lambda_{\rm fs} &=\int_{a_d}^{a_{eq}} \frac{da'}{a'} \frac{1}{a' H(a')} \left(\frac{v_{\rm kick}~a_d}{a'}\right)\\
        &=\int_{a_d}^{a_{eq}} \frac{da'}{a'^3}\frac{1}{H_0 \sqrt{\Omega_r a'^{-4}}} v_{\rm kick}~a_d\\
        &= \frac{v_{\rm kick}~(\Gamma_\phi)^{-1}}{a_d} \ln \left(\frac{a_{eq}}{a_d}\right)
    \end{split}
\end{equation}
where, in the last line, we noted that the decay happens when $\Gamma_\phi = H(a_d)$.

%\begin{widetext}
    \begin{table*}[ht]
    \centering
    \caption{Benchmark Points Comparison}
        \begin{tabular}{|l|c|c|c|c|c|c|c|c|c|}
        \hline%\toprule
         & $~m_{\phi} (\text{GeV})~$ & $~m_{\chi}(\text{GeV})~$ & $y$ & $~v_{\rm kick}~$ & $a_d$ & $a_{nr}$ & $a_{\rm eq}$ & $\lambda_{\rm fs} (\text{Mpc})$ & $~\Delta N_{\rm eff}~$\\
         \hline
        \hline%\midrule
        $\textbf{BP1}$    & 22.4   & 2.82   & $~~1.8\times10^{-17}$ & ~0.968~ & $~~5.2\times10^{-4}$ & $~~1.76\times10^{-3}$ & $~2.94\times10^{-4}$ & 0.256 & 0.13 \\
        \hline%\midrule
        $\textbf{BP2}$    & 46.56   & 1   & $~~8.1\times10^{-16}$ & ~0.999~ & $~~6.84\times10^{-6}$ & $~~1.56\times10^{-4}$ & $~2.94\times10^{-4}$ & 0.006 & 0.023\\
        \hline%\midrule
        $\textbf{BP3}$   & 127.32   & 0.1178  & $~~8.75\times10^{-15}$ & ~0.999~ & $~~3.83\times10^{-7}$ & $~~2.07\times10^{-4}$ & $~2.94\times10^{-4}$ & 0.0005 & 0.04\\
        \hline\bottomrule
    \end{tabular}
    \label{tab:benchmarks}
\end{table*}
%\end{widetext}

We summarize three representative benchmark scenarios in Table~\ref{tab:benchmarks}, illustrating that, with an appropriate choice of parameters, it is indeed possible to satisfy the structure formation constraints.
 In Fig.~\ref{fig:FSL}, we present the parameter space points that simultaneously satisfy the observed dark matter relic abundance and yield $\Delta N_{\rm eff}$ in the range $[0.01, 0.2]$, while ensuring the free-streaming constraint $\lambda_{\rm fs} < 0.1$ Mpc is respected. The plot displays $\lambda_{\rm fs}$ as a function of $m_\phi$, with the color bar representing variations in the Yukawa coupling $y$. The DM mass is automatically adjusted so that the combined thermal and non-thermal contributions yield the correct relic density as shown in Fig.~\ref{fig:freezeINfraction}.

% We generated a plot of the $\lambda_{\rm fs}$ using points that give rise to the correct dark matter relic density
% as well as the $\Delta N_{\text{eff}}$ which lies in the region $[0.01-0.2]$. The Fig~\ref{fig:FSL} is plotted as a function of $m_\phi$, with the colour gradient showing the variation of the Yukawa coupling $y$.

% Moreover, dark matter, which constitutes a significant portion of the total energy budget of the Universe, is governed by $\Omega_{\rm CDM}h^2 = 0.12 \pm 0.001$ and consists of two key components: the freeze-out component, which originates from the thermal bath, and the freeze-in component, which arises from late-time $\phi$ decay.  

% Examining the freeze-in contribution shown in Fig.~\ref{fig:freezeINfraction}, it is evident that if the $\phi$ particle is significantly more massive than $\chi$, the fraction of total DM originating from $\phi$ decay increases. Moreover, a large mass difference between the parent and daughter particles results in a highly relativistic kick velocity for DM (see Fig.~\ref{fig:kick_velocity}) when produced near the CMB epoch. To prevent this, $m_\phi$ must be kept close to $m_\chi$, thereby reducing the kick velocity and ensuring that DM remains non-relativistic, without violating observational constraints from Planck data.  

\section{Summary and Conclusion}\label{sec::conclusion}

In this work, we have proposed and analyzed a minimal yet phenomenologically rich framework for light self-interacting dark matter within a gauged $U(1)_D$ extension of the Standard Model. This setup not only addresses longstanding small-scale structure anomalies in the $\Lambda$CDM paradigm but also predicts observable signatures in the effective number of relativistic species, $\Delta N_{\rm eff}$. By combining thermal freeze-out with late time non-thermal production from scalar decays, the framework reconciles the relic abundance of light dark matter with large self interaction cross-sections capable of resolving small scale issues. Simultaneously, the model generates a detectable $\Delta N_{\rm eff}$ through dark radiation produced during scalar decays, offering complementary cosmological signatures.

The dark sector in our model consists of a fermionic dark matter candidate $\chi$, a light mediator $X_\mu$, a scalar $\phi$, and a sterile neutrino $\nu_S$. The relic abundance of $\chi$ is achieved through a hybrid mechanism involving both thermal freeze-out and non-thermal production from the late decay of $\phi$, which occurs after BBN but before the CMB epoch. This decay not only replenishes the dark matter density but also produces dark radiation $\nu_S$, contributing to $\Delta N_{\rm eff}$, thus offering an additional observational signature. Similar phenomenology can be achieved even if we allow $\phi$ decay to DM and active neutrinos facilitated via active-sterile mixing similar to seesaw models of neutrino mass. Our analysis demonstrates that this setup naturally accommodates the required strong self-interactions for $\chi$, ensuring compatibility with astrophysical observations on small scales, while simultaneously satisfying constraints from relic abundance, direct detection experiments, and structure formation constraint on the free-streaming length. We also showed that the model can give rise to a measurable $\Delta N_{\rm eff}$, lying within the projected sensitivity reach of future CMB experiments like SPT-3G, CMB-S4, and CMB-HD.

In summary, this framework presents a coherent and testable scenario that links the solution to small-scale structure anomalies with observable cosmological signatures. Future precision measurements in both cosmology and direct detection experiments will play a decisive role in probing and potentially validating this scenario.

\acknowledgments
The work of D.B. is supported by the Science and Engineering Research Board (SERB), Government of India grants MTR/2022/000575 and CRG/2022/000603. D.B. also acknowledges the support from the Fulbright-Nehru Academic and Professional Excellence Award 2024-25. S.M. acknowledges the financial support from the National Research Foundation (NRF) grant funded by the Korea government (MEST) NRF-2022R1A2C1005050. The work of N.S. is supported by the Department of Atomic Energy-Board of Research in Nuclear Sciences, Government of India (Grant No. 58/14/15/2021- BRNS/37220).

 \section*{}
 \appendix\label{appendix}

\section{Stueckelberg Mechanism}\label{app:stueckelberg}
The relevant part of the Lagrangian of the Stueckelberg extension of the SM is given by \cite{Aboubrahim:2022bzk,Kors:2004dx}
\begin{equation}
    \mathcal{L} \supset -\frac{1}{4} C^{\mu \nu}C_{\mu \nu}-\frac{1}{2}\left( M_1 C_\mu + M_2 B_\mu + \partial_\mu \sigma \right)^2.
\end{equation}
Here, $C_\mu$ is the gauge field of the $U(1)_D$, and $B_\mu$ is that for the $U(1)_Y$. The field $\sigma$ is an axionic field which will be absorbed by the $C_\mu$ in the unitary gauge, thereby giving mass term for the $C_\mu$ field. The term $M_2$ which does not appear in the the SM Lagrangian is explicitly written in the Stueckelberg extension to explain the Stueckelberg mass mixing \cite{Kors:2004dx}. However, we consider a simpler model without the Stueckelberg mass mixing, and hence we will take $M_2$ to be simply 0. Then the Lagrangian above can be made gauge invariant by splitting off the longitudinal degrees of freedom of $\sigma$ through 
\begin{equation}
    \begin{split}
        \hat{X_\mu} = & ~C_\mu +\frac{1}{M_1}\partial_\mu \sigma\\
    \end{split}
    \label{eq:stueckelberg}
\end{equation}
along with the variations in the fields
\begin{equation}
        \delta C_\mu = \partial_\mu \alpha, \,\, \delta \sigma =-M_1 \alpha
\end{equation}
where $\alpha$ is a gauge parameter. Then, the final Lagrangian for a massive gauge boson without violating the gauge structure of the theory takes the form
\begin{equation}
    \mathcal{L} \supset -\frac{1}{4} \hat{X}^{\mu \nu}\hat{X}_{\mu \nu}-\frac{1}{2} M_X^2 \hat{X^\mu} \hat{X_\mu}.
    \label{eq:Stueckelberg_term}
\end{equation}.

\section{X boson interaction with SM fermions}\label{app:fields}
Various interactions induced due to the $\epsilon$ mixing.\\
\textbf{Electromagnetic current: $j_{\rm em}^\mu$}
\begin{equation}
    \begin{split}
        \hat{A}_\mu j_{\rm em}^\mu & = A'_\mu j_{\rm em}^\mu + \frac{-\epsilon~cw}{\sqrt{1-\epsilon^2}}X'_\mu j^{\mu}_{\rm em}\\
        & = A_\mu j_{\rm em}^\mu\\
        & ~ - \frac{\epsilon~cw}{\sqrt{1-\epsilon^2}}(\cos\eta X_\mu - \sin\eta Z_\mu)j_{\rm em}^\mu.
    \end{split}
\end{equation}
\\
\textbf{Neutral current: $j_Z^\mu$}
\begin{equation}
    \begin{split}
        \hat{Z}_\mu j_{Z}^\mu & = Z'_\mu j_Z^\mu + \frac{\epsilon~sw}{\sqrt{1-\epsilon^2}}X'_\mu j_Z^\mu\\
        & =(\cos\eta Z_\mu + \sin\eta X_\mu)j_Z^\mu\\
        & ~~+ \frac{\epsilon~sw}{\sqrt{1-\epsilon^2}}~(\cos\eta X_\mu - \sin\eta Z_\mu)j_Z^\mu\\
        & \simeq (\sin\eta  + \cos\eta \frac{\epsilon~sw}{\sqrt{1-\epsilon^2}})X_\mu j_Z^\mu\\
        & ~~+ \cos\eta Z_\mu j_Z^\mu.
    \end{split}
\end{equation}
\\
Now combining the result of the $X_\mu$ interaction with the SM $j^\mu_{Z}$ and $j^\mu_{\rm em}$, we have the final interaction terms given by
\begin{equation}
    \begin{split}
        X_\mu j^\mu_{SM} &= X_\mu
        \left( \sin\eta + \cos\eta \frac{\epsilon~sw}{\sqrt{1-\epsilon^2}} \right) j_{Z}^\mu \\
        & ~~- X_\mu \left(\cos\eta \frac{\epsilon~cw}{\sqrt{1-\epsilon^2}} \right) j_{\rm em}^\mu.
    \end{split}
\end{equation}

\section{Low energy cross-sections relevant for the self-interactions of dark matter}\label{app::sidm_cross-section}
Since DM self-interaction is realized due to the presence of the term like $g_{\chi} X_\mu \overline{\chi}\gamma^\mu \chi$, the non-relativistic DM self-scattering can be well understood in terms of the attractive Yukawa potential
 \begin{equation}
     V(r)=\pm \frac{g_{\chi}^2}{4\pi r}e^{-M_{X}r}
 \end{equation}
where the $+(-)$ sign denotes repulsive (attractive) potential.\\
\\
To capture the relevant physics of forward scattering, the transfer cross-section is defined as
 \begin{equation*}
     \sigma_T=\int d\Omega (1-cos\theta)\frac{d\sigma}{d\Omega}
 \end{equation*}
 In the Born limit, ${g_{\chi}}^2 m_{\chi}/(4\pi {M_{X}})\ll 1$,

\begin{equation*}
     \sigma^{\rm Born}_T=\frac{{g_{\chi}}^4}{2\pi {m_{\chi}^2} v_{\rm rel}^4} \left[ \log\left( 1+\frac{m^2_{\chi} v_{\rm rel}^2}{M_{X}^2} \right)-\frac{m_{\chi}^2 v_{\rm rel}^2}{M_{X}^2+m_{\chi}^2 v_{\rm rel}^2} \right]
\end{equation*}

Outside the Born limit, where   ${g_{\chi}}^2 m_{\chi}/(4\pi {M_{X}})\geq 1$, there can be two different regions: classical regime and resonance regime. In the classical regime (${m_{\chi}v_{\rm rel}}/{M_{X}}\geq 1$), solution for an attractive potential is given by \cite{Tulin:2012wi, Tulin:2013teo, PhysRevLett.90.225002}

\begin{equation*}
   \sigma^{\rm class.}_T=\begin{cases}
     \frac{4\pi}{M_{X}^2}\beta^2 \ln(1+\beta^{-1}) & \textbf{$\beta > 1$}\\
     \frac{8\pi}{M_{X}^2}\left[ {\beta^2/(1+1.5\beta^{1.65})} \right] & \textbf{$10^{-1} < \beta \leq 10^3$} \\
     \frac{\pi}{M_{X}^2}\left[ \ln\beta +1-1/2 \ln^{-1}{\beta}\right]^2 & \textbf{$\beta \geq 10^3$}
   \end{cases}
\end{equation*}
and for the repulsive potential
\begin{equation*}
 	\sigma^{\rm class.}_T=\begin{cases}
 		\frac{2\pi}{M_{X}^2}\beta^2 \ln(1+\beta^{-2}) & \textbf{$\beta \leq 1$}\\
 		\frac{\pi}{M_{X}^2}\left[ {{\rm ln}(2 \beta) -{\rm ln}({\rm ln}2 \beta)} \right]^2 & \textbf{$ \beta \geq 1$} 
 	\end{cases}
\end{equation*}
where $\beta=\frac{2g_{\chi}^2{M_{X}}}{4\pi {m_{\chi}}v_{\rm rel}^2}$. 

Finally in the resonance region (${m_{\chi}v_{\rm rel}}/{M_{X}}\leq 1$), no analytical formula for $\sigma_T$ is available. So approximating the Yukawa potential by Hulthen potential $\left(V(r)=\pm \frac{{g_{\chi}}^2}{4\pi}\frac{\delta e^{-\delta r}}{1-e^{-\delta r}}\right)$, the transfer cross-section is obtained to be: 

\begin{equation*}
         \sigma_T^{\rm Hulthen}=\frac{16\pi \sin^2\delta_0}{m^2_\chi v_{\rm rel}^2}
\end{equation*}
 where $l=0$ phase shift $\delta_0$ is given by:
 $$ \delta_0=Arg \left[ \frac{i\Gamma(im_{\chi}v_{\rm rel}/\kappa M_{X})}{\Gamma(\lambda_+)\Gamma(\lambda_-)} \right]$$
 {\rm with}
 $$\lambda_{\pm}=1+\frac{im_{\chi}v_{\rm rel}}{2 \kappa M_{X}}\pm i^r\sqrt{\frac{{g_{\chi}}^2 m_{\chi}}{4\pi \kappa M_{X}}+(-1)^{r+1}\frac{m_{\chi}^2{v_{\rm rel}^2}}{4 \kappa^2 {M^2_{X}}}}$$
 
 and $\kappa \approx 1.6$ is a dimensionless number and $r=0(1)$ for attractive(repulsive) potential.

% {\color{red} Lets not use v for velocity, use some subscript like $v_\chi, v_{\rm rel}$ as in main text. v is used for vev earlier. Also, lets not use $a$ for abother parameter as it is already used for scale factor.}
\section{Thermally Averaged Cross-section}\label{app::th_avergae}
\begin{figure}[h]
	\centering
    \includegraphics[width=0.15\textwidth]{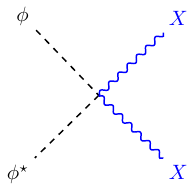}
    \includegraphics[width=0.23\textwidth]{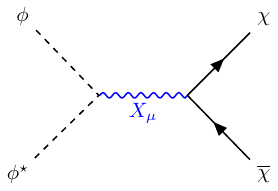}
	\caption{Feynman diagrams for the number changing processes, relevant for the calculation of $\phi$ and $\chi$ abundance.}
	\label{fig:phi_annihilation}
\end{figure}

\begin{figure}[h]
	\centering
	\includegraphics[width=0.3\textwidth]{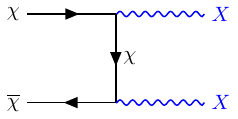}
	\caption{Feynman diagram of $\chi$ annihilation to $X$.}
	\label{fig:dm_annihilation}
\end{figure}

The cross-section for the process $\phi \phi \rightarrow X X$ (Fig.~\ref{fig:phi_annihilation}) and that of $\chi \chi \rightarrow X X$ (Fig.~\ref{fig:dm_annihilation}) are given by
\begin{widetext}
    \begin{equation*}
    \begin{split}
        \sigma_{\phi \rightarrow X}&=\frac{g_d^4}{8 \pi s (s-4m_\phi^2)}\times \left( \frac{\sqrt{(s-4m_\phi^2)(s-4M_X^2)}\left(21 M_X^4+16 m_\phi^4+m_\phi^2(20 s-88 M_X^2)\right)}{2(M_X^4+m_\phi^2 (s-4M_X^2))} \right) \\
        &~~\left.-\frac{(2s^2+4s~M_X^2-24(s-M_X^2)m_\phi^2+16m_\phi^4-7M_X^4)}{(s-2 M_X^2)} \text{Log}\left(\frac{2M_X^2-s+\sqrt{(s-4m_\phi^2)(s-4M_X^2)}}{2M_X^2-s-\sqrt{(s-4m_\phi^2)(s-4M_X^2}}\right)\right),  \\
        \sigma_{\chi \rightarrow X}&=\frac{g_d^4}{4 \pi s (s-4m_\phi^2)}\times \left( -\frac{\sqrt{(s-4m_\chi^2)(s-4M_X^2)}\left(s~m_\chi^2+4 m_\chi^4-2 M_X^4\right)}{2(M_X^4+m_\chi^2 (s-4M_X^2))} \right. \\
        &~~-\left.\frac{(s^2+4(s-2M_X^2)m_\chi^2-8m_\chi^4+4M_X^4)}{2(s-2 M_X^2)}\text{Log}\left(\frac{2M_X^2-s+\sqrt{(s-4m_\chi^2)(s-4M_X^2)}}{2M_X^2-s-\sqrt{(s-4m_\chi^2)(s-4M_X^2}}\right)\right).
    \end{split}
\end{equation*}
\end{widetext}

To compute the analytic expression of the thermally averaged cross-section suitable for the thermal freeze-out scenario, we write $s=4 m_{\phi}^2+m_{\phi}^2 \text{v}^2$, and write $\sigma v$ in powers of $\text{v}$, as follows
\begin{eqnarray}
    \sigma_{\phi \rightarrow X} & = & \frac{g_d^4 \sqrt{-M_X^2+m_\phi^2}}{16 \pi \text{v} m_\phi^3 (2m_\phi^2-M_X^2)^2}\left(24m_\phi^4-24m_\phi^2 M_X^2+7M_X^4 \right) \nonumber \\
    & &\quad \quad +\mathcal{O(\text{v})} \nonumber
\end{eqnarray}
Ignoring the higher order terms in v, we get
\begin{equation*}
    \begin{split}
        \sigma_{\phi \rightarrow X} & \approx \frac{g_d^4}{16 \pi m_\phi^3 \text{v}}\sqrt{-M_X^2+m_{\phi}^2} \frac{24m_\phi^4-24m_\phi^2 M_X^2+7M_X^4}{(2m_\phi^2-M_X^2)^2}\\
        &\approx \frac{g_d^4}{16 \pi m_\phi^2 \text{v}}\sqrt{1-\frac{M_X^2}{m_{\phi}^2}} \frac{24 m_\phi^4}{(2 m_\phi^2)^2}\\
        &= 6 \frac{g_d^4}{16 \pi m_\phi^2 \text{v}}\sqrt{1-\frac{M_X^2}{m_{\phi}^2}}
    \end{split}
\end{equation*}

where, in the last two line, we have used the condition that $M_X<m_\phi$. Then, $ \sigma \text{v} $ becomes
\begin{equation}
    \left.\sigma \text{v}\right|_{\phi \rightarrow X} = 6\times \frac{g_d^4}{16 \pi m_\phi^2}\sqrt{1-\frac{M_X^2}{m_\phi^2}}.
\end{equation}
% \begin{figure}[h]
% 	\centering
%     \includegraphics[width=0.3\textwidth]{Images/thermal_average.png}
% 	\caption{The thermally averaged cross-section for the various processes. N denotes the numerical solution, and A denotes the analytic solution.}
% 	\label{fig:th_average}
%\end{figure}
Similarly, for $\chi \chi \leftrightarrow XX$, the corss-section can be expanded in powers of v using $s=4 m_{\chi}^2+m_{\chi}^2 \text{v}^2$ as 
\begin{equation*}
    \begin{split}
        \sigma_{\chi \rightarrow X} & = \frac{g_d^4 (-M_X^2+m_\chi^2)^{3/2}}{4 \pi m_\chi \text{v} (2m_\chi^2-M_X^2)^2}~+\mathcal{O(\text{v})}\\
        &\approx \frac{g_d^4}{4 \pi m_\chi \text{v}}\sqrt{-M_X^2+m_{\chi}^2}\times \frac{(m_\chi^2-M_X^2)}{(2m_\chi^2-M_X^2)^2}\\
        &\approx \frac{g_d^4}{4 \pi \text{v}}\sqrt{1-\frac{M_X^2}{m_{\chi}^2}}\times \frac{m_\chi^2}{(2 m_\chi^2)^2}\\
        &= \frac{g_d^4}{16 \pi m_\chi^2 \text{v}}\sqrt{1-\frac{M_X^2}{m_{\chi}^2}}
    \end{split}
\end{equation*}
where, again, we ignored the higher power of v and realizing that $M_X<m_\chi$. Then, $\sigma \text{v}$ for the $\chi$ annihilation becomes 
\begin{equation}
    \left.\sigma \text{v}\right|_{\chi \rightarrow X}= \frac{g_d^4}{16 \pi m_\chi^2}\sqrt{1-\frac{M_X^2}{m_\chi^2}}.
    \label{eq:dm_annihilation}
\end{equation}
% The thermally averaged annihilation cross-sections for all the processes are shown in Fig.~\ref{fig:th_average}.

\section{Interaction rate of scattering processes}\label{app::int_rate_scattering}

To find the thermally-averaged cross-section for a process $1+2 \leftrightarrow 3+4$, we use the general formula \cite{Gondolo:1990dk}
\begin{equation}
    \langle \sigma v \rangle = \frac{\int \sigma v e^{-E_1/T} e^{-E_2/T} d^3p_1 d^3p_2}{\int e^{-E_1/T} e^{-E_2/T} d^3p_1 d^3p_2}.
    \label{eq:thermal_average}
\end{equation}
Rewriting the integration measure as 
\begin{equation*}
    \begin{split}
        d^3p_1 d^3p_2 &= \left( 4\pi |p_1| E_1 dE_1 \right) \left( 4\pi |p_2| E_2 dE_2 \right) \frac{1}{2} d(cos(\rm \theta))\\
        &= (2\pi^2) E_1 E_2 dE_+dE_- ds,
    \end{split}
\end{equation*}
where the 2nd step is achieved via change of variables; $E_+=E_1+E_2$, $E_-=E_1-E_2$ and $s=E_1^2+E_2^2+2E_1 E_2 -2 |p_1||p_2|cos(\rm \theta)$. Noting the region of integration as
\begin{equation}
    \begin{split}
        s &\geq (m_1+m_2)^2\\
        E_+ &\geq \sqrt{s}\\
        \left| E_-E_+ +\frac{m_2^2-m_1^2}{s} \right| &\leq 2p_{12}\sqrt{\frac{E_+^2-s}{s}}
    \end{split}
\end{equation}
The numerator (N) of Eq.~\eqref{eq:thermal_average} can be simplified by performing a first integration on $dE_-$ as $\int dE_- =\frac{4 p_{12}}{\sqrt{s}}\sqrt{E_+^2-s}$ to get the following
\begin{equation}
    \begin{split}
        \rm N &= 2\pi^2 \int ds dE_+ \sigma v E_1 E_2 \times \frac{4 p_{12}}{\sqrt{s}}\sqrt{E_+^2-s}~e^{-E_+/T}\\
        &= 2\pi^2 \int ds dE_+ \left(p_{12} \sqrt{s} \sigma \times \frac{4 p_{12}}{\sqrt{s}}\right)\sqrt{E_+^2-s}~e^{-E_+/T}\\
        &= 8\pi^2 \int ds~\sigma (s) p_{12}^2(s) \times \int dE_+ e^{-E_+/T} \sqrt{E_+^2-s}
    \end{split}
\end{equation}
where we have used $\sigma v E_1 E_2=p_{12}(s) \sqrt{s}~\sigma$ \cite{Edsjo:1997bg} to get the 2nd line for $p_{12}$ defined by $p_{12}(s)=\left([s-(m_1+m_2)^2]^{1/2}~[s-(m_1-m_2)^2]^{1/2}\right)/2 \sqrt{s}$. Finally, integrating w.r.t. $dE_+$, the numerator becomes 
\begin{equation}
    N = 8\pi^2 T \int ds~\sigma (s) p_{12}^2(s) \sqrt{s} K_1 \left( \sqrt{s}/T \right).
\end{equation}
The denominator of Eq.~\eqref{eq:thermal_average} can be simplified as 
\begin{equation}
    \begin{split}
        D & = \int d^3p_1 d^3p_2 f_1f_2\\
        & =\frac{(2 \pi)^6}{g_1 g_2} \times \int g_1\frac{d^3p_1 f_1}{(2 \pi)^3} \times \int g_2\frac{d^3p_2 f_2}{(2 \pi)^3}\\
        & = \frac{(2 \pi)^6}{g_1 g_2} n_1n_2.
    \end{split}
\end{equation}
Thus, the thermally-averaged cross-section for the scattering becomes
\begin{equation}
    \begin{split}
        \langle \sigma v \rangle = &\frac{g_1 g_2}{(2 \pi)^6 n_1 n_2} 8\pi^2 T \\
        & \times \int^{\infty}_{(m_1+m_2)^2}ds~ \sigma (s) p_{12}^2(s) \sqrt{s} K_1 \left( \sqrt{s}/T \right).
    \end{split}
\end{equation}
and the interaction rate is $n_{bath} \times \langle \sigma v \rangle$.
% Finally, the interaction rate
% for the process $\phi b \leftrightarrow \phi b$ can be 
% calculated as $n_b \times \langle \sigma v \rangle_{\phi b \leftrightarrow \phi b}$.

% \begin{equation}
%     \begin{split}
%         \Gamma_{\phi}& 
%         % & =n_b \times \frac{g_\phi g_b}{(2 \pi)^6 n_\phi n_b} 8\pi^2 T \\
%         % & ~~\times \int^{\infty}_{(m_\phi+m_b)^2}ds~ \sigma (s) p_{12}^2(s) \sqrt{s} K_1 \left( \sqrt{s}/T \right)\\
%          = \frac{g_b}{4 \pi^2 m_\phi^2 K_2 \left( \frac{m_\phi}{T} \right)}\\
%         & ~~~~~\times \int^{\infty}_{(m_\phi+m_b)^2}ds~ \sigma (s) p_{12}^2(s) \sqrt{s} K_1 \left( \frac{\sqrt{s}}{T} \right).
%     \end{split}
% \end{equation}

\section{Complete Boltzmann Equations}\label{app::complete_BE}
The complete Boltzmann equations concerning the evolution of $X$, $\phi$, and $\chi$ are given by
{\begin{widetext}
\begin{equation}
    \begin{split}
        \frac{dY_{X}}{dx} = & \frac{s(m_{\phi})}{H(m_{\phi})} \frac{\langle \sigma v \rangle_{\phi \rightarrow X}}{x^2}\left[Y_{\phi}^2(x)- \frac{Y_X^2}{(Y^{eq}_X(x~M_X/m_\phi))^2} (Y^{eq}_{\phi})^2(x) \right] \\
        & +\frac{s(m_{\phi})}{H(m_{\phi})} \frac{\langle \sigma v \rangle_{\chi \rightarrow X}}{x^2}\left[Y_{\chi}^2(x)- \frac{Y_X^2}{(Y^{eq}_X(x~M_X/m_\phi))^2}  (Y^{eq}_{\chi})^2(x~m_\chi/m_\phi) \right] - \frac{x~\Gamma_{X \rightarrow SM}}{H(m_{\phi})} \left[ Y_{X}(x) - Y^{eq}_X(x~M_X/m_\phi)\right],\\
        \frac{dY_{\chi}}{dx} = & -\frac{s(m_{\phi})}{H(m_{\phi})} \frac{\langle \sigma v \rangle_{\chi \rightarrow X}}{x^2}\left[Y_{\chi}^2(x)- \frac{Y_X^2}{(Y^{eq}_X(x~M_X/m_\phi))^2}(Y^{eq}_{\chi})^2(x~m_\chi /m_\phi) \right] ~ + \frac{x~\Gamma_{\phi}}{H(m_{\phi})} Y_{\phi}(x),\\
        \frac{dY_{\phi}}{dx} = & ~- \frac{s(m_{\phi})}{H(m_{\phi})} \frac{\langle \sigma v \rangle_{\phi \rightarrow X}}{x^2}\left[Y_{\phi}^2(x)- \frac{Y_X^2}{(Y^{eq}_X(x~M_X/m_\phi))^2}(Y^{eq}_{\phi})^2(x)\right]~ - \frac{x~\Gamma_{\phi}}{H(m_{\phi})} Y_{\phi}(x),
    \end{split}
    \label{eq:abundance0}
\end{equation}
\end{widetext}
where $x=m_{\phi}/T$ is the dimensionless parameter, $s(m_\phi)$ and $H(m_\phi)$ are the corresponding entropy density and the Hubble expansion rate expressed in terms of $m_\phi$ (with $x$ factored out). 
% Solution to the Eqn.~\ref{eq:abundance} is plotted in Fig.~\ref{fig:complete_BE}.
% \begin{figure}[h]
% 	\centering
%     \includegraphics[width=0.23\textwidth]{Images/complete_BE_relic.png}
%     \includegraphics[width=0.23\textwidth]{Images/complete_BE_energyD.png}
% 	\caption{\textbf{Left}: Evolution of the co-moving number densities of $X$, $\phi$ and $\chi$ , illustrating the determination of the dark matter relic density. In the absence of \(X\), a choice of \(m_\phi = 20\) GeV results in the correct relic abundance. \textbf{Right}: Evolution of the new relativistic species, showing a shift in $\Delta N_{\rm eff}$ from 0.15 to 0.17 when the equation for $X$ is switched off.}
% 	\label{fig:complete_BE}
% \end{figure}
For most calculations, we adopt a simplified form of the Boltzmann equation that does not require tracking the co-moving number density of $X$.

\section{Decay Width of X}\label{app::decay_width}
The interaction given in Eq. \eqref{eq:X-interaction} can be used to calculate the decay width of X to electron and neutrino. They are calculated as 
\begin{equation}
    \Gamma_{X \rightarrow \nu\overline{\nu}}= \frac{(C^\nu_L)^2 M_X}{24 \pi},
\end{equation}
\begin{equation}
    \Gamma_{X \rightarrow e^+e^-}= \frac{(C^e_{LR})^2 M_X}{24 \pi} \sqrt{1-\frac{4 m_e^2}{M_X^2}},
\end{equation}
where the factor $(C^e_{LR})^2$ is given by
\begin{equation}
\left( (C^e_L)^2+(C^e_R)^2 \right) \left(1-\frac{m_e^2}{M_X^2}\right) + 6 \frac{m_e^2}{M_X^2} C^e_L C^e_R.
\end{equation}

	\bibliographystyle{apsrev}
	\bibstyle{apsrev}
%	\bibliography{ref,refn,ref1,ref12,ref3,ref4,ref5}	

\end{document}